\documentclass[prl,aps,twocolumn,superscriptaddress]{revtex4-1}

\usepackage{graphicx,color}
\usepackage{amsthm}
\usepackage{amsfonts}
\usepackage{algorithmic}
\usepackage{enumerate}
\usepackage{latexsym}
\usepackage{amsmath}
\usepackage{amssymb}
\usepackage{ulem}
\usepackage{color}
\usepackage[colorlinks=true,citecolor=blue,linkcolor=blue]{hyperref}

\def\avg#1{\left\langle#1\right\rangle}

\emergencystretch=\maxdimen
\include{MyCommand}

\begin{document}
\title{Enhancement of $d$-wave pairing in the striped phase with the nearest neighbour attraction}
\author{Lufeng Zhang}
\affiliation{School of Science, Beijing University of Posts and Telecommunications, Beijing 100876, China\\}
\author{Ting Guo}
\affiliation{Department of Physics, Beijing Normal University, Beijing
100875, China\\}
\author{Yingping Mou}
\affiliation{Beijing Computational Science Research Center, Beijing
100193, China}
\author{Qiaoni Chen}
\affiliation{Department of Physics, Beijing Normal University, Beijing
100875, China\\}
\author{Tianxing Ma}
\email{txma@bnu.edu.cn}
\affiliation{Department of Physics, Beijing Normal University, Beijing
100875, China\\}
\affiliation{Beijing Computational Science Research Center, Beijing
100193, China}

\begin{abstract}
Recent experimental results of the angle-resolved photoemission spectroscopy suggested that, an additional strong nearest neighbor attraction in the Hubbard model might be significant for the doped cuprates. The stripe-ordered patterns, which are formed by the inhomogeneous distribution of the spin, charge and pairing correlations in the CuO$_{2}$ planes, are a well-known feature of the doped cuprates. By employing the constrained path quantum Monte Carlo, we examine the effect of the nearest-neighbor attraction on the two-dimensional repulsive Hubbard model. Within and across the stripe regions, the ground state spin correlations and the $d$-wave pairing correlation are calculated. We found that the spin-spin correlation is the largest when the inter-stripe region is nearly half filled, and we also found the $d$-wave superconducting correlation of the neighboring sites is enhanced in the presence of the stripe pattern. This reveals the crucial effects of the strong nearest-neighbor attraction on superconductivity in doped cuprates.

\end{abstract}

\pacs{71.10.Fd, 74.20.Rp, 74.70.Xa, 75.40.Mg}

\maketitle

\noindent

\underline{\it Introduction} --- In 1986, Bednorz and Muller discovered high-temperature superconductivity (HTSC) in the Ba-La-Cu-O system\cite{1986Possible}.
Soon after this discovery, extensive research was conducted to raise the critical temperature $T_{c}$ of HTSC in doped cuprates and explore the superconducting mechanism\cite{RevModPhys.78.17,ANDERSON1196}.
Since the Ba-La-Cu-O system is a strongly correlated material,  it is common to introduce nearly degenerate state close to the superconducting region\cite{Dagotto2005} in order to obtain novel phases\cite{Dagotto2005}. The stripe-ordered patterns, are a well-known feature of doped cuprates; they are formed by the inhomogeneous distribution of the spin, charge and pairing correlations in the CuO$_{2}$ planes\cite{1995Evidence,RevModPhys.75.1201,PhysRevB.40.7391,MACHIDA1989192,doi:10.1143/JPSJ.59.1047}. Numerous studies have attempted to ascertain the inhomogeneous order of the spin and charge stripes, paired-waves, and unconventional superconductivity\cite{PhysRevLett.88.167008,2005Spatially,2020Atomic,2016Chuanzhou}. The understanding of how these orders compete and cooperate with each other to produce exotic phases, and how they induce the high-temperature superconducting phase of doped cuprates is the most significant problem in condensed matter physics for several decades\cite{RevModPhys.87.457,nature14165}.

The two-dimensional  Fermionic Hubbard model\cite{PhysRevB.39.9749,PhysRevLett.64.1445,PhysRevB.42.10641,PhysRevB.42.6809,PhysRevB.43.11442,
PhysRevB.62.12700,PhysRevB.54.R8281,Gehlhoff_1996} and $t-J$ model\cite{PhysRevLett.64.475,PhysRevLett.78.4609,PhysRevB.62.13930,PhysRevB.62.1684,
PhysRevB.57.627,PhysRevB.63.014414} are widely used to characterize the phase separation and superconductivity in doped cuprates. Many properties of the cuprate superconductors are well described by the Hubbard model, such as the Mott insulating phase, the suppression of the antiferromagnetic order upon doping, and stripe formation. As superconductivity appears to compete or coexist with the stripe formation, simulations are required to describe all phases and reveal the possible superconducting phase, which approaches the zero temperature and thermodynamic limit. Although the simple Hubbard model can describe the basic dispersion structure, it does not accurately address the additional spectral features\cite{doi:10.1126/science.abf5174}.
The results of recent angle-resolved photoemission spectroscopy (ARPES) experiments indicate that the nearest-neighbor attractive Coulomb interaction $V$ could enhance the spectral weight of the holon folding branch of doped one-dimensional (1D) cuprate chains\cite{doi:10.1126/science.abf5174}. A comparison of the momentum distribution curves (MDCs) obtained from ARPES showed, that the highest agreement with a doping-independent $V$ was between $-1.2t$ and $-0.8t$, and the spectral intensities were accurately predicted by the 1D Hubbard model.
Due to structural and quantum chemistry similarities among cuprates, the nearest neighbor attraction effect should not be ignored, including $d$-wave
superconductivity in two-dimensional systems. In this work, we analyze the additional Coulomb interactions in the two-dimensional repulsive Hubbard model to examine the relationship between superconductivity and the stripe patterns of doped cuprates.


In the doped situation, which is away from half filling, the phase of cuprates could be separated into half-filled antiferromagnetic regions and hole concentrated areas.
Another theoretical framework of phase separation is the striped phase, which was proposed theoretically in 1990s\cite{PhysRevB.40.7391,MACHIDA1989192,doi:10.1143/JPSJ.59.1047} and first observed by Tranquada et al. experimentally in 1995\cite{1995Evidence}.
Static stripes are widely found in insulators, and there is a lot of strong evidence for the existence of dynamic fringes in metals and superconducting compounds\cite{PhysRevB.59.14712,PhysRevB.60.3643,PhysRevLett.118.177601}.
Incommensurate spin fluctuations, which may originate from the dynamic fringe phase, were observed in the La$_{1.6-x}$Nd$_{0.4}$Sr$_{x}$CuO$_{4}$ system\cite{PhysRevLett.67.1791}.
Tranquada et al. used Nd to replace part of the La to introduce lattice distortion and ``freeze" the dynamic stripes into the static state.
Then they observed the charge and spin stripes using elastic neutron scattering\cite{1995Evidence}.
Yamada et al. studied the La$_{2-x}$Sr$_{x}$CuO$_{4}$ system without Nd in more detail, and the results strongly indicate the existence of the dynamic fringe phase\cite{PhysRevB.57.6165}.
There are many more experimental evidences that there is a close relationship between the stripe phase and high-$T_{c}$ superconductivity\cite{1995Evidence,2005Spatially,doi:10.1126/sciadv.aax3346}.
High-$T_{c}$ superconducting copper oxides all have layered perovskite structures, and studies have shown that the striped phase can effectively inhibit the Josephson coupling between layers\cite{PhysRevLett.99.127003}. Therefore, we can obtain the basic properties of high-$T_{c}$ superconductors and the relationship between high-$T_{c}$ superconductivity and the striped phase by analyzing the two-dimensional extended Hubbard model.

Unlike in 1D problems, the analytic solutions of quantum many-body Hamiltonian in higher dimension are quite rare, so two dimensional and higher dimensional numerical simulations are essential in order to compare with realistic systems. The quantum Monte Carlo simulations are one of the most widely used type of numerical methods. Early numerical results obtained from the determinant quantum Monte Carlo (DQMC) method indicated, that the extended $s$-wave \cite{PhysRevB.91.241107} and $d$-wave\cite{PhysRevB.39.839,PhysRevB.37.5070} pairing susceptibilities\cite{RevModPhys.66.763} were dominant.
Unfortunately, the DQMC method has the limitations of finite temperatures and system sizes due to the infamous fermion sign problem.
Thus, it is difficult to identify the existence of the superconducting state using the DQMC simulations\cite{PhysRevB.90.075121,PhysRevB.91.241107}.

In this paper, we utilize the constrained path quantum Monte Carlo(CPQMC) method to study the stripe pattern in a two-dimensional Hubbard model on a square lattice\cite{PhysRevB.55.7464,PhysRevLett.74.3652}.
The CPQMC method has been used as a benchmarking tool to calculate the ground state energy and other observables in various systems\cite{PhysRevB.84.121410,HUANG2019310,PhysRevB.101.155413}.
Another reason to choose the CPQMC method is that it prevents the infamous sign problem\cite{PhysRevB.41.9301} encountered in the DQMC method when dealing with systems that are far from being half-filled. For example, the sign problem exists even at temperatures of $\beta=6$ with a wide doping range of $0.6<\rho<1.0$\cite{PhysRevB.86.184506,li2021dopingdependent}.
The basic strategy of CQPMC is to project out the ground-state wave function from an initial wave function by branching random walk in an overcomplete space of constrained Slater determinants, which have positive overlaps with a known
trial wave function. In this work, we focus on the closed-shell case, for which the corresponding free-electron wave function is non-degenerate and translationally invariant. In this case, the free-electron wave function is a good choice as the trial wave function. For more details about the CQPMC method are described in the Appendix.We investigate three aspects: the relationship between the striped phase and the density of the particles, the spin correlation function and its influencing factors, and the $d$-wave pairing correlation function and its influencing factors.

\begin{figure}
\includegraphics[scale=0.55]{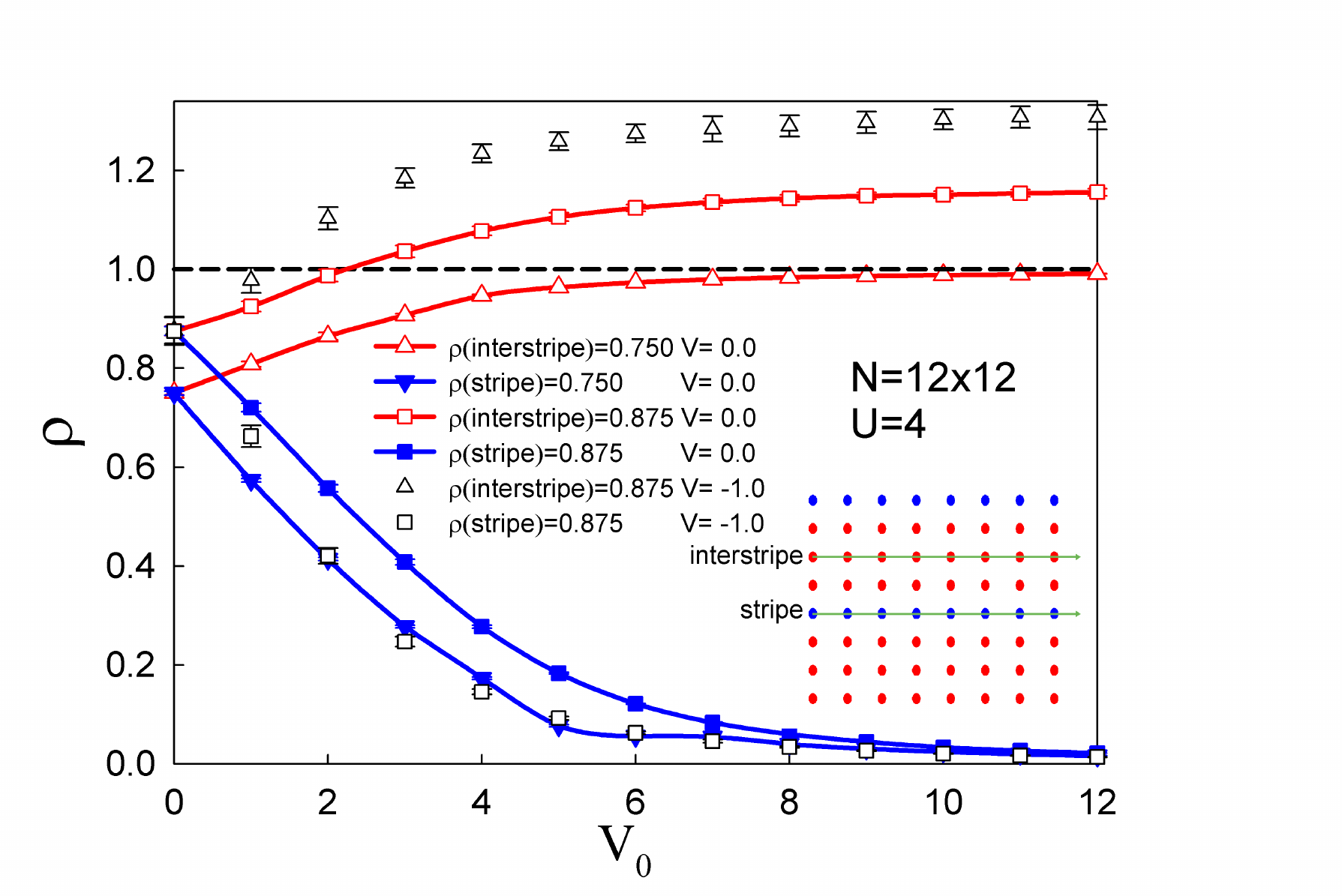}
\caption{The density of particles on and in-between the stripes as a function of $V_{0}$ when the total density is fixed. The total density of the lattice is fixed at $\rho=0.750$ and $\rho=0.875$. The lattice size is $12\times12$, $U=4$. Blue filled symbols represent the striped rows and red empty symbols represent the unstriped rows. The inset in the left bottom is the sketch of $8\times8$ square lattice with stripe period $P=4$, where the blue colors label the site with stripe potential $V_{0}$, and the red dots represent sites without $V_{0}$.}
\label{Fig:density}
\end{figure}

\noindent
\underline{\it Model and numerical method}
We consider the two-dimensional repulsive Hubbard Hamiltonian on a square lattice. Stripes are introduced externally via $V_{0}$ on a set of rows with a period $P=4$. The diagram is shown on the bottom left in Fig.~\ref{Fig:density}.
Thus, the Hubbard Hamiltonian is written as
\begin{align}
H = & -\sum_{\langle \bf{i,j}\rangle\sigma}
       t^{\phantom{\dagger}}
       (c^{\dagger}_{\bf{i}\sigma}c^{\phantom{\dagger}}_{\bf{j}\sigma} +c^{\dagger}_{\bf{j}\sigma}c^{\phantom{\dagger}}_{\bf{i}\sigma})
      +U\sum_{\bf{i}} n_{\bf{i}\uparrow}n_{\bf{i}\downarrow}\nonumber\\
      &-\mu\sum_{\bf i}(n_{\bf i \uparrow} + n_{\bf i \downarrow}) +V\sum_{\langle \bf{i,j}\rangle} n_{\bf{i}\uparrow}n_{\bf{j}\downarrow} +V_{0}\sum_{\bf{i}_{y}\in P}\left(n_{\bf{i}\uparrow}+n_{\bf{i}\downarrow}\right).
   \label{eq:model}
\end{align}
Here, $c^{\dagger}_{\bf{i}\sigma}(c^{\phantom{\dagger}}_{\bf{i}\sigma})$ are creates(annihilates) operators acting at site $\bf{i}$, and $n_{\bf{i}\sigma}=c^{\dagger}_{\bf{i}\sigma}c^{\phantom{\dagger}}_{\bf{i}\sigma}$ is the occupy number operator. We only consider the hopping kinetic energy $t$ between the nearest neighbor lattice sites $\langle \bf{i,j}\rangle$. $U$ is the Coulomb repulsion representing the energy consumed by the double occupation of electrons on the same lattice, while $V$ represents the interaction between the nearest neighbors. $V_{0}$ is an additional on-site energy imposed on a set of rows $\bf{i}=(\bf{i}_{x}+\bf{i}_{y})$ with $mod(\bf{i}_{y},P)=0$. $V_{0}$ has the same roles as $\mu$ but acts locally. Since the interest in the charge order patterns in HTSC is related to the correlated quasi-1D or quasi-2D electronic structures\cite{1995Stripe,PhysRevLett.78.338,1975Charge,PhysRevB.37.562,Pierre2012}, we set $P=4$ as the periodic number of the stripe charge patterns,
which means $V_{0}$ is imposed every four rows.

In order to investigate the size effects, the simulations we performed are mainly on three different lattice sizes: $8\times8$, $12\times12$ and $16\times16$,  The inset of Fig.~\ref{Fig:density} shows, the lattices of $L=8$ with $P=4$. The blue dots represent the sites where the $V_{0}$ term is active, whereas the red dots represent the sites where the $V_{0}$ term is inactive. The entire $8\times8$, $12\times12$ and $16\times16$ lattice accommodate two, three and four stripes respectively in the $P=4$ case. Our data present in this work are mainly performed on the average total density $\rho=\avg{n}= 0.875$. At this filling, it allows for the existence of a broad range of densities on the stripe and between stripes, and also the charge order is strongest at doping $1/8$ in cuprates according to the experimental results\cite{Ghiringhelli821,PhysRevB.90.054513,He608}.


We focus on two physical quantities in this paper. One is the $C_{spin}$, spin correlation function, which reflects the spin distribution,
\begin{align}
 C_{spin}(\bf{i}) =&
   \langle S^{-}_{\bf{j}+\bf{i}}S^{+}_{\bf{j}}\rangle \nonumber\\
 S^{+}_{\bf{j}}=&c^{\dagger}_{\bf{j}\uparrow}c^{\phantom{\dagger}}_{\bf{j}\downarrow}
 \label{eq:condform}
\end{align}
Here $S^{+}_{\bf{j}}$ is the spin at site $\bf{j}$. If $C_{spin}(\bf{i})>0$, the spin direction at $\bf{i}$ site is the same as the spin direction at $\bf{j}$ site. If $C_{spin}(\bf{i})<0$, the spin direction at $\bf{i}$ site is opposite to the spin direction at $\bf{j}$ site. We can determine the magnetic strength of the system and the type of magnetism, such as antiferromagnetism, ferromagnetism, or long-range order.

Another quantity is the $d$-wave pairing correlation function $P^{d}$. The superconducting phase of doped cuprates comes from the electron pair, which is the Bose condensation of Cooper pairs at low temperature. It differs from the $s$-wave pair of conventional superconductors, it is a $d$-wave pair.
We investigate factors affecting the $d$-wave pairing by analyzing the pairing correlation function, which is written as
\begin{align}\label{eq:Saf}
P^{d}(\bf{i}) & = \langle \Delta_{d}({\bf i} + {\bf j}) \Delta^{\dagger}_{d}({\bf j}) \rangle \nonumber\\
\Delta^{\dagger}_{d} (\bf{j}) &= c^{\dagger}_{\bf{j}\uparrow}(c^{\dagger}_{\bf{j}+\hat{x}\downarrow}-c^{\dagger}_{\bf{j}+\hat{y}\downarrow}
+c^{\dagger}_{\bf{j}-\hat{x}\downarrow}-c^{\dagger}_{\bf{j}-\hat{y}\downarrow})
\end{align}

We use the CPQMC method, which was first benchmarked and described in detail by Zhang\cite{PhysRevLett.74.3652,PhysRevLett.84.2550}.
Its basic calculation principle is $|\varphi_{g}\rangle=\lim\limits_{\beta\rightarrow\infty}e^{-\beta\hat{H}}|\varphi_{T}\rangle$.
The two key concepts of the CPQMC method are importance sampling and constrained path approximation. Importance sampling is used to evaluate the importance of the sampling variables and increase the sampling opportunities of variables with a greater impact on the system to improve the iterative efficiency. The Monte Carlo method can be regarded as a random walk process with multiple independent samples. It has the advantages of high efficiency and fast convergence. However, the random walk causes some problems, such as the sign problem in quantum Monte Carlo simulations of fermion systems. The CPQMC method has no sign problem since it uses the constrained path approximation. Therefore, the CPQMC method can deal with many systems that cannot be analyzed by the conventional quantum Monte Carlo method. It accurately predicts the symmetry of the magnetic and superconducting pairing using the two-dimensional Hubbard model\cite{PhysRevLett.84.2550}.  We have provided additional information on the CPQMC method in the Appendix. More technical details on the CPQMC method can be found in Refs\cite{PhysRevLett.74.3652,PhysRevB.55.7464,PhysRevB.64.205101}.

\noindent
\underline{\it Results and discussion} --- We show in Fig.~\ref{Fig:density}, that the density of particles on and in-between the stripes changes with $V_{0}$ when the total density is fixed. The figure shows the result for the $8\times8$ lattice with $\rho=0.750$ and $\rho=0.875$. When $V_{0}=0$, the densities on the stripes and interstripes are equal; thus the lattice is homogeneous. As $V_{0}$ increases, the density on the stripes decreases, and the density on the interstripes increases. The result indicates that $V_{0}$ ensures that electrons flow from the stripe region to the interstripe region. It should be noted that the interstripe region reaches the half-filling state at $V_{0}\approx3$ and $V_{0}\approx8$ for $\rho=0.875$ and $\rho=0.750$ respectively. According to the Pauli exclusion principle, a lattice can contain a maximum of two electrons at most. A lattice containing one electron, is called a half-filled lattice. Once the interstripes reaches half-filling, the densities on and in-between the stripe are almost unchanged.

\begin{figure}
\includegraphics[scale=0.28]{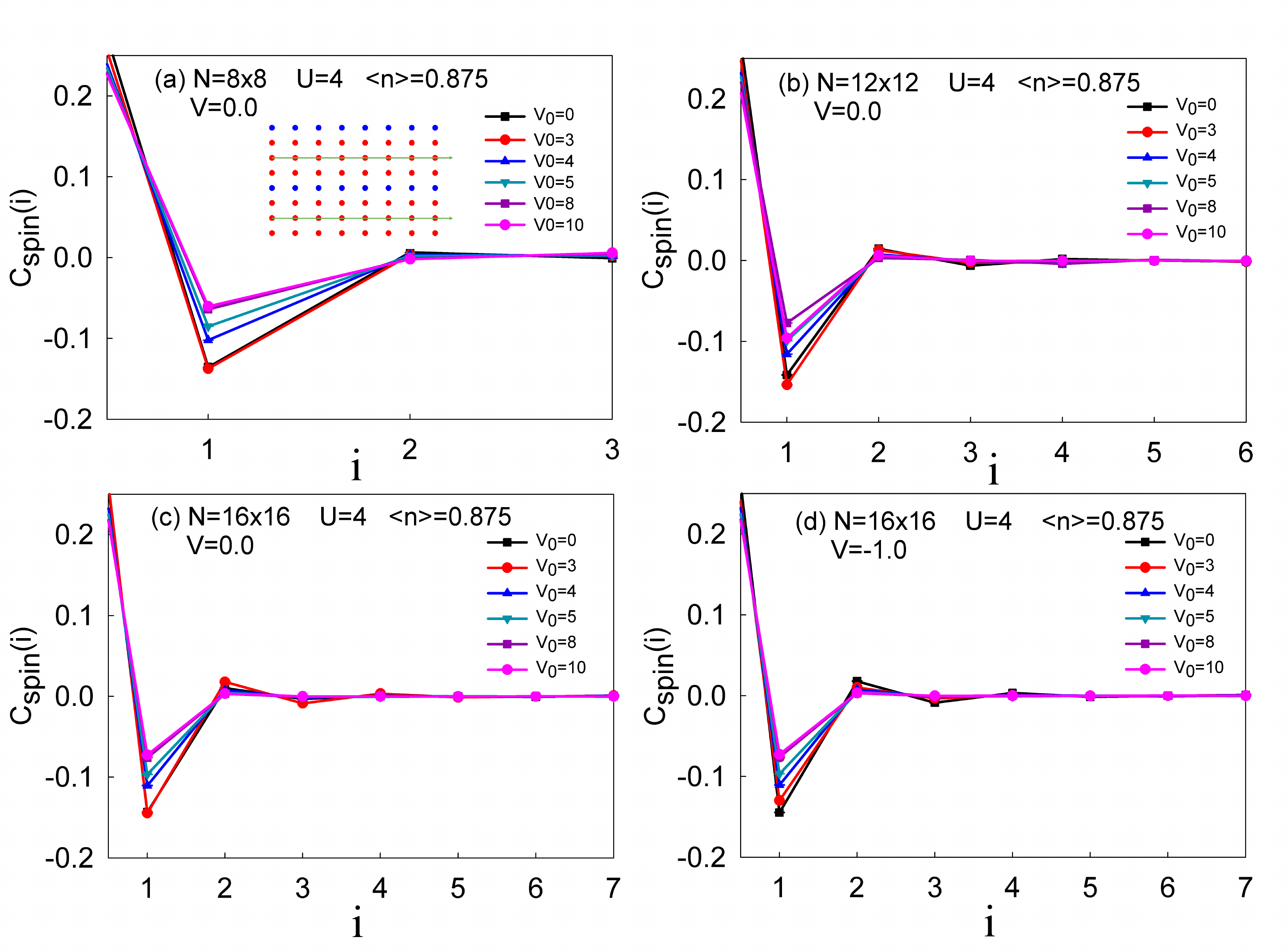}
\caption{The spin correlation function $C_{spin}(\bf{i})$ under different $V_{0}$ along the center of interstriped rows. The total density is fixed at $\rho=0.875$. The lattice size is (a) $8\times8$, (b) $12\times12$ and (c) $16\times16$. The results with the nearest neighbor attractive coulomb interaction are shown in panel (d).}
\label{Fig:Cspin}
\end{figure}

\begin{figure}
\includegraphics[scale=0.28]{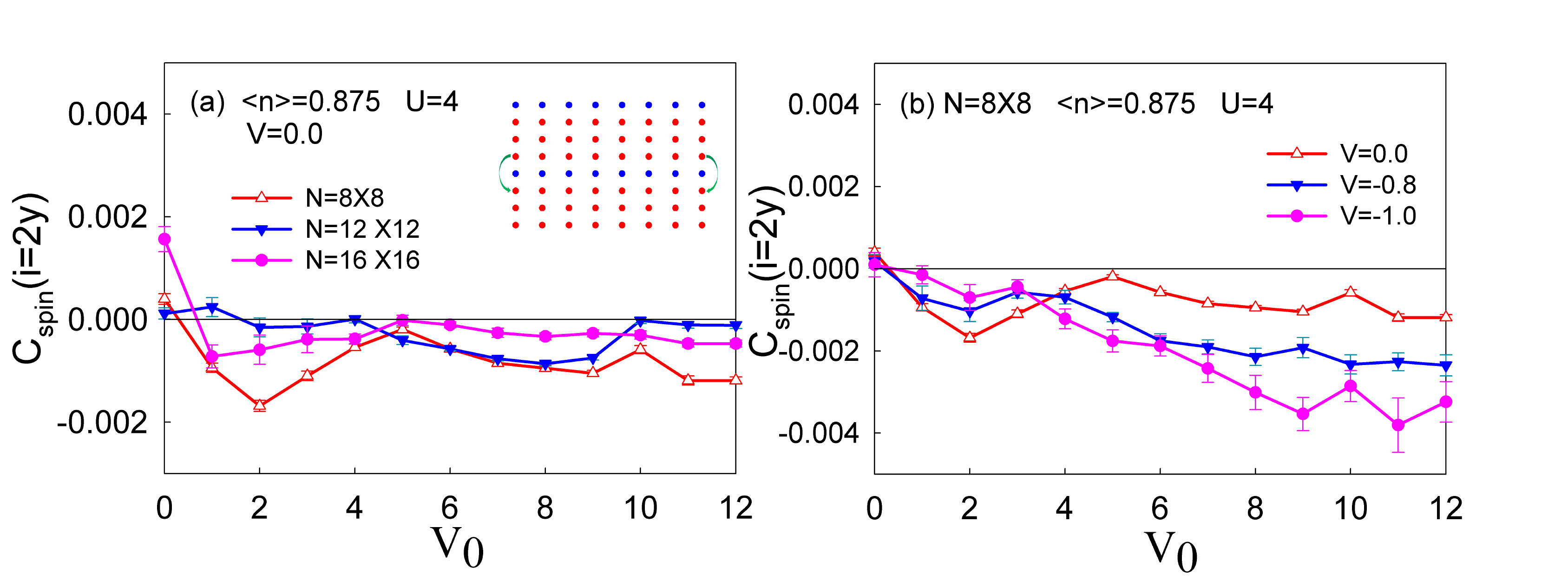}
\caption{(a) The spin correlation function $C_{spin}({\bf{i}}=2y)$ as a function of $V_{0}$. In this figure, we show the spin correlation function for fixed distance ${\bf{i}}=2y$ which crosses a stripe. (b) $C_{spin}({\bf{i}}=2y)$ behavior in the case of the nearest neighbor attractive coulomb interaction $V=-0.8$ and $V=-1.0$. The total density of the lattice is fixed at $\rho=0.875$.}
\label{Fig:Cspin_cross}
\end{figure}

In Fig.~\ref{Fig:Cspin}, the spin correlation function $C_{spin}({\bf{i}})$ is shown at different $V_{0}$ for $\rho=0.875$ along the center of interstriped rows, as shown by the arrow in the inset of Fig.~\ref{Fig:Cspin}.
We can see that antiferromagnetic correlations are short ranged for all $V_{0}$ and the spin correlations are small. As presented in Fig.~\ref{Fig:Cspin} (a)-(c), without the nearest coulomb interactions $V$, the spin correlation strength is maximum for all three different size lattices around $V_{0}=3$. For $V_{0}=3$ and $\rho=0.875$, the interstriped region is half-filling which suggests that half-filling is beneficial to the formation of spin correlations and antiferromagnetic correlations. While in Fig.~\ref{Fig:Cspin} (d), the attractive coulomb interaction is presented, $V=-1.0$, the spin correlation decrease gradually as the stripe potential $V_0$ grows.

Besides being interested in the spin correlation function along the interstriped rows, we are also interested in the spin correlation function which crosses a stripe.
As mentioned above, with the increasement of $V_{0}$, the density on stripes decreases. Because the low density can not have a large moment, the spin correlations should be reduced with $V_{0}$.
In Fig.~\ref{Fig:Cspin_cross} (a), we show the $C_{spin}({\bf{i}}=2y)$ as a function of $V_{0}$. $C_{spin}({\bf{i}}=2y)$ reflects the spin correlation between a pair of sites traversing a stripe, as shown by the arrows in Fig.~\ref{Fig:Cspin_cross}(a). Besides, we concern the sign of $C_{spin}({\bf{i}}=2y)$. As we can see in panel (a) of Fig.~\ref{Fig:Cspin_cross}, $C_{spin}({\bf{i}}=2y)$ is negative for small $V_{0}$. However, for $V_{0}>1$, $C_{spin}({\bf{i}}=2y)$ turns negative for $8\times8$ lattice. For $V_{0}=0$, the sign of $C_{spin}({\bf{i}}=2y)$ is positive.
This is the $\pi$-phase shift of spin which is a prominent experimental feature of stripe physics in the cuprates\cite{1995Stripe}.
This result shows that it can be observed on two-dimensional model with stripes. Further more, we also checked the case with the nearest neighbor coulomb interaction as Fig.~\ref{Fig:Cspin_cross} (b) exhibited.
The $\pi$-shift behavior is enhanced by the presence of $V$.

\begin{figure}
\includegraphics[scale=0.28]{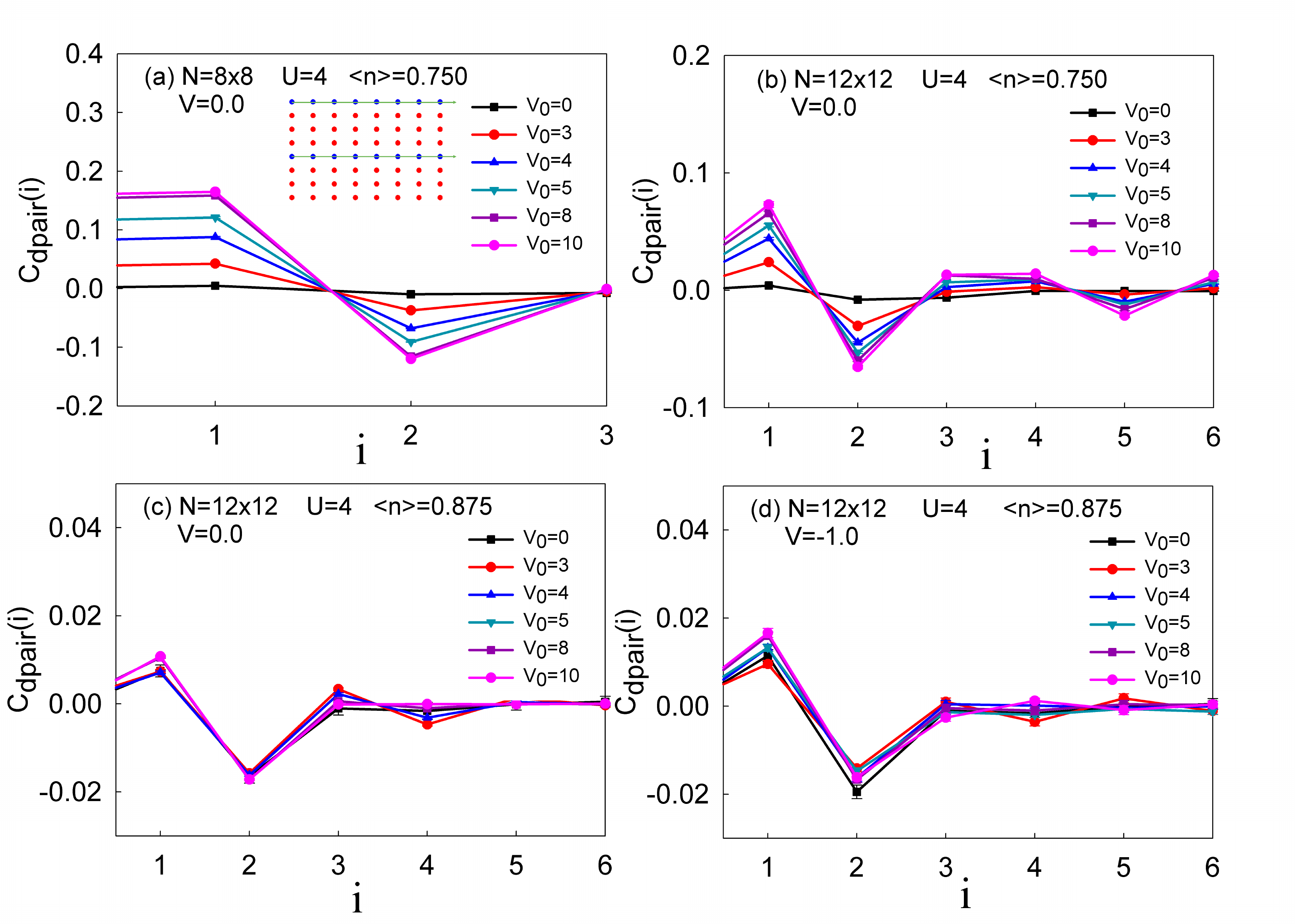}
\caption{
The $d$-wave pairing function $C_{dpair}(\bf{i})$ under different $V_{0}$ along the striped rows. The total density of the lattice is discussed on $\rho=0.875$ and $\rho=0.750$. The lattice size is (a) $8\times8$, $\rho=0.750$, (b)$ 12\times12$, $\rho=0.750$ and (c) $12\times12$, $\rho=0.875$. (d) $C_{dpair}(\bf{i})$ behavior in consideration of near-neighbor attractive coulomb interaction $V=-1.0$.}
\label{Fig:Cdpair}
\end{figure}

\begin{figure}
\includegraphics[scale=0.28]{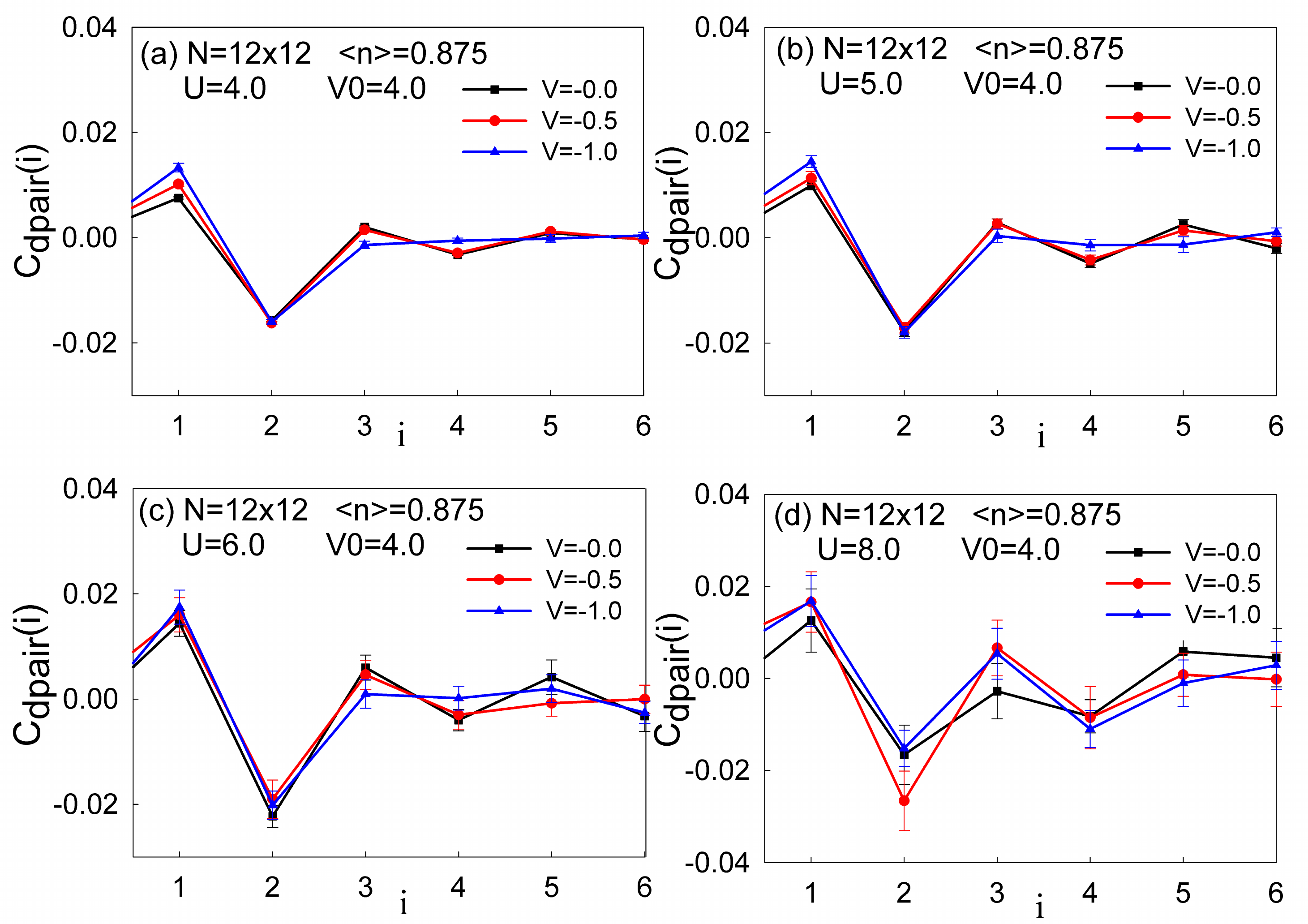}
\caption{The d-wave pairing function $C_{dpair}(i)$ under different near-neighbor attractive coulomb interaction $V=0,-0.5,-1.0$ along the stripe rows. The total density of the lattice is fixed at $\rho=0.875$. The on-site coulomb strengths are (a) $U=4.0$, (b) $U=5.0$, (c) $U=6.0$ and (d) $U=8.0$. }
\label{Fig:CdpairU4568}
\end{figure}

Then we analyze the $d$-wave pairing correlation function. In Fig.~\ref{Fig:Cdpair}, we show the $d$-wave pairing correlation function under different $V_{0}$ along the striped rows, as shown by the arrows in Fig.~\ref{Fig:Cdpair}(a).
The analysis shows that there is almost no $d$-wave pairing between the other lattices except the nearest neighbor.
The main reason that we choose $U=4$ is that the coulomb interaction is large enough for correlated systems, and the simulations could be very difficult with greater $U$. In order to solve this problem further, we extended our simulations to other $U$ values, such as $U=4,5,6,8$ as present in Fig.~\ref{Fig:CdpairU4568}. According to the results, the $d$-wave pairing function exhibit similar behaviors with different $U$ strengths. As the near-neighbor interaction enhanced, the strength of $d$-wave paring get greater. Thus, we could rely on our simulation results present in the manuscript which are mainly calculated on $U=4$, the physics behind is clear enough.

\begin{figure}
\includegraphics[scale=0.28]{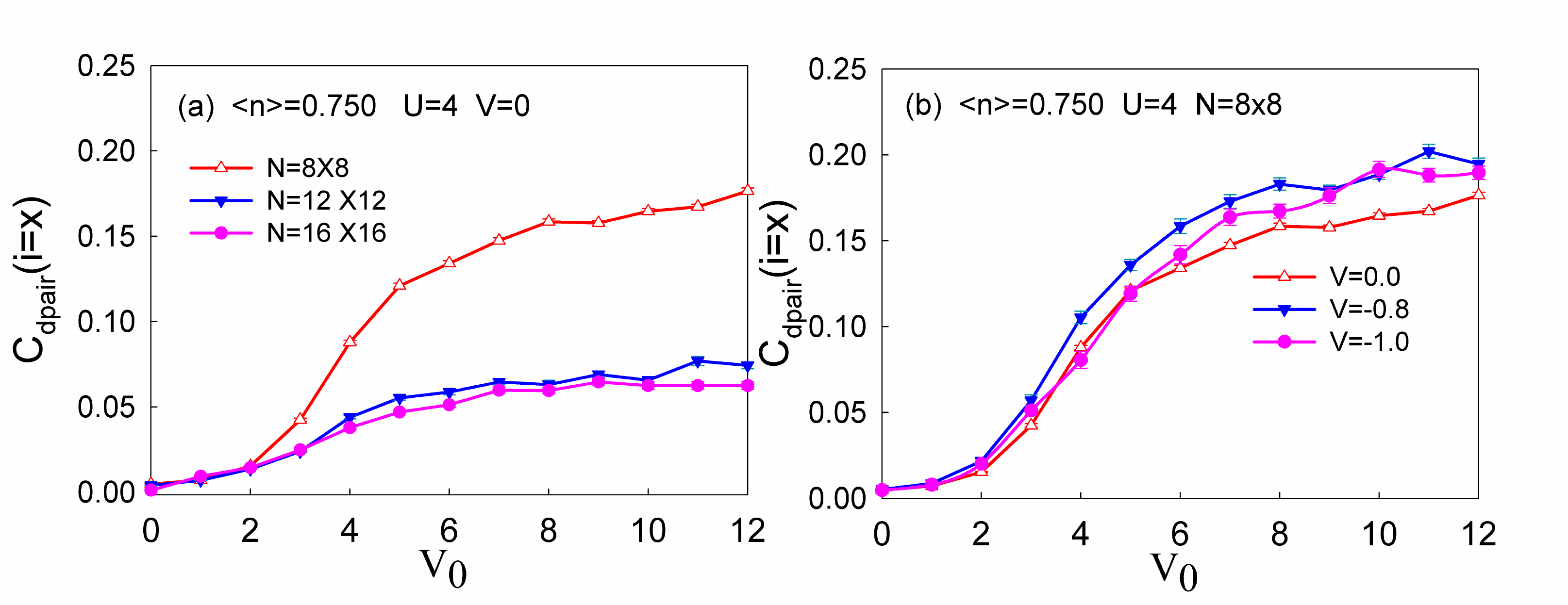}
\caption{The $d$-wave pairing function $C_{dpair}(\bf{i}=x)$ on neighboring sites along the stripes as a function of $V_{0}$. (a) $C_{dpair}(\bf{i}=x)$ behavior on different lattice sizes. (b)
In the case of near-neighbor attractive coulomb interaction $V=-0.8$ and $V=-1.0$. The total density of the lattice is fixed at $\rho=0.750$.
}
\label{Fig:Cdpair_nn}
\end{figure}

The $d$-wave pairing correlation function is short ranged. Therefore, we focus on the analysis of the $d$-wave pairing correlation function between the nearest neighboring sites.
As shown in Fig.~\ref{Fig:Cdpair_nn}(a), we study the $d$-wave pairing correlation function $C_{dpair}({\bf{i}}=x)$ on neighboring sites along stripes as a function of $V_{0}$.
It is significant that the $d$-wave pairing correlation function is enhanced by the application of $V_{0}$.
The larger $V_{0}$ is, the stronger $d$-wave pairing is.
The superconducting phase of copper oxide high-$T_{c}$ superconductor comes from $d$-wave electron pairing.
The above results show that one can enhance the $d$-wave pairing between the nearest neighbor lattice points by applying $V_{0}$, imposing charge stripes on materials.
In panel (b) of Fig.~\ref{Fig:Cdpair_nn}, the effect of the nearest neighbor attractive interaction is discussed.
As the data illustrated, the $d$-wave pairing strength is enhanced by $V$.
It indicates that the nearest neighbor attractive interaction could also intensify the $d$-wave pairing pattern.

\noindent
\underline{\it Summary} ---
The existence of the stripe pattern and the additional nearest neighbor attractive interactions causes a significant enhancement of the $d$-wave paring in the doped two-dimensional repulsive Hubbard Hamiltonian model. The stripe order is introduced by inhomogeneous charge distributions at period $P=4$ suggested by the neutron scattering\cite{1995Evidence}. Based on the two-dimensional Hubbard model, the imposing stripe patterns $V_{0}$ makes electrons flow from stripe region to interstripe region. When the lattice is half-filled, the spin correlation is the strongest. Half-filling is beneficial to the formation of spin correlations and antiferromagnetic correlations.

The enhancement of $d$-wave pairing is also observed in our results. With the increasement of additional stripe potential $V_0$ at $P=4$, the $d$-wave pairing correlation function is significantly enhanced. Spontaneously, in the presence of the nearest neighbor attractive interaction, the $d$-wave pairing pattern becomes more robust. Stripe phase, doping and additional near-neighbor attractive interaction, we combined these three effects on Hubbard model, drawing a conclusion that they all play crucial roles in driving significant enhancement on $d$-wave pairing. It would be interesting to explore the possible enhancement of pairing by other types of charge inhomogeneities and at other doping levels.

\noindent
\underline{\it Acknowledgement} ---
This work was supported by NSFC (No. 11774033 and 11974049) and Beijing Natural Science Foundation (No. 1192011). The numerical simulations in this work were performed at HSCC of Beijing Normal University and Tianhe in the Beijing Computational Science Research Center.

%
%
%
%
%
%
%
%
%
\begin{center}
\textbf{Appendix: Constrained Path Monte Carlo method}
\end{center}
\setcounter{equation}{0}

Our calculations were performed on the square lattices of $N=L\times L$ unit cells with periodic boundary conditions imposed using the CQPMC method. The basic strategy of CQPMC is to project out the ground-state wave function $|\psi
_{0}\rangle $ from an initial wave function $|\psi
_{\mathcal{T}}\rangle $ by branching random walk in an overcomplete space of constrained Slater determinants $%
|\phi \rangle $, which have positive overlaps with a known trial wave function. In this work, we start with the Hamiltonian,
\begin{align}
H = & -\sum_{\langle \bf{i,j}\rangle\sigma}
       t^{\phantom{\dagger}}
       (c^{\dagger}_{\bf{i}\sigma}c^{\phantom{\dagger}}_{\bf{j}\sigma} +c^{\dagger}_{\bf{j}\sigma}c^{\phantom{\dagger}}_{\bf{i}\sigma})
      +U\sum_{\bf{i}} n_{\bf{i}\uparrow}n_{\bf{i}\downarrow}\nonumber\\
      &-\mu\sum_{\bf i}(n_{\bf i \uparrow} + n_{\bf i \downarrow}) +V\sum_{\langle \bf{i,j}\rangle} n_{\bf{i}\uparrow}n_{\bf{j}\downarrow} +V_{0}\sum_{\bf{i}_{y}\in P}\left(n_{\bf{i}\uparrow}+n_{\bf{i}\downarrow}\right).
   \label{eq:model}
\end{align}
We project out the
ground state by iterating
\begin{equation}
|\psi ^{\prime }\rangle =e^{-\Delta \tau (H-E_{T})}|\psi \rangle
\end{equation}%
where $E_{\mathcal{T}}$ is some guess of the ground-state energy.
Purposely $\Delta \tau $ is a small parameter so for
$H=\mathcal{T}+\mathcal{V }$ we can write $e^{-\Delta\tau H}\approx e^{-\Delta\tau \mathcal{T}/2}e^{-\Delta\tau \mathcal{V}}e^{-\Delta\tau \mathcal{T}/2}$ through the first-order Trotter approximation. $\mathcal{T}$ and $\mathcal{V}$ are the kinetic and potential energy operators.

Because the kinetic energy is a quadratic form in the creation and
destruction operators for each spin, the action of its exponential on the
trial state is simply to transform one direct product of Slater determinants
into another. While the potential energy is not a quadratic form in the
creation and destruction operators, its exponential is replaced by the sum of
exponentials of such forms via the discrete Hubbard-Stratonovich
transformation. For the on-site Coulomb term, this transformation is
\begin{equation}
e^{-\Delta \tau Un_{i,\sigma }n_{i,-\sigma }}=\frac{1}{2}\sum_{x=\pm
1}e^{x\Delta \tau J(n_{i,\sigma }-n_{i,-\sigma })}e^{-\frac{1}{2}\Delta \tau
U(n_{i,\sigma }+n_{i,-\sigma })}
\end{equation}%
Here, $U\geq0$ and $cos(i\Delta\tau J)=e^{\Delta\tau U/2}.$ For the nearest neighbor Coulomb repulsion term, we make the same type of
transformation but we have to do it many more times: $n_{i}n_{j}=n_{i%
\uparrow }n_{j\uparrow }+n_{i\uparrow }n_{j\downarrow
}+n_{i\downarrow }n_{j\uparrow }+n_{i\downarrow }n_{j\downarrow }$.

One consequence of the Hubbard-Stratonvich transformation is the
factorization of the projection into an up and down spin part. Accordingly
we re-express the iteration step of Eq. (2) as
\begin{equation}
\prod_\sigma |\phi_\sigma^{\prime }\rangle = \int d\vec x\, P(\vec x)
\prod_\sigma B_\sigma(\vec x)|\phi_\sigma\rangle
\end{equation}
where $\vec x =(x_1,x_2,\dots,x_N)$ is the set of Hubbard-Stratonovich
fields (one for each lattice site), $N$ is the number of lattice sites, $%
P(\vec x)=(\frac{1}{2})^N$ is the probability distribution for these fields,
and $B_\sigma(\vec x)$ is an operator function of these fields formed from
the product of the exponentials of the kinetic and potential energies.

The Monte Carlo method is used to perform the multi-dimensional
integration over the Hubbard-Stratonovich fields. It does so by
generating a set of random walkers initialized by replicating
$|\psi_\mathcal{T}\rangle$ many times. Each walker is then
propagated independently by sampling a $\vec x$ from $P(\vec x)$ and
propagating it with $B(\vec x)$. After the propagation has
``equilibrated'', the sum over the walkers provides an estimate of
the ground-state wave function $|\psi_0\rangle$.

\bibliography{reference}

\begin{thebibliography}{60}%
\makeatletter
\providecommand \@ifxundefined [1]{%
 \@ifx{#1\undefined}
}%
\providecommand \@ifnum [1]{%
 \ifnum #1\expandafter \@firstoftwo
 \else \expandafter \@secondoftwo
 \fi
}%
\providecommand \@ifx [1]{%
 \ifx #1\expandafter \@firstoftwo
 \else \expandafter \@secondoftwo
 \fi
}%
\providecommand \natexlab [1]{#1}%
\providecommand \enquote  [1]{``#1''}%
\providecommand \bibnamefont  [1]{#1}%
\providecommand \bibfnamefont [1]{#1}%
\providecommand \citenamefont [1]{#1}%
\providecommand \href@noop [0]{\@secondoftwo}%
\providecommand \href [0]{\begingroup \@sanitize@url \@href}%
\providecommand \@href[1]{\@@startlink{#1}\@@href}%
\providecommand \@@href[1]{\endgroup#1\@@endlink}%
\providecommand \@sanitize@url [0]{\catcode `\\12\catcode `\$12\catcode
  `\&12\catcode `\#12\catcode `\^12\catcode `\_12\catcode `\%12\relax}%
\providecommand \@@startlink[1]{}%
\providecommand \@@endlink[0]{}%
\providecommand \url  [0]{\begingroup\@sanitize@url \@url }%
\providecommand \@url [1]{\endgroup\@href {#1}{\urlprefix }}%
\providecommand \urlprefix  [0]{URL }%
\providecommand \Eprint [0]{\href }%
\providecommand \doibase [0]{http://dx.doi.org/}%
\providecommand \selectlanguage [0]{\@gobble}%
\providecommand \bibinfo  [0]{\@secondoftwo}%
\providecommand \bibfield  [0]{\@secondoftwo}%
\providecommand \translation [1]{[#1]}%
\providecommand \BibitemOpen [0]{}%
\providecommand \bibitemStop [0]{}%
\providecommand \bibitemNoStop [0]{.\EOS\space}%
\providecommand \EOS [0]{\spacefactor3000\relax}%
\providecommand \BibitemShut  [1]{\csname bibitem#1\endcsname}%
\let\auto@bib@innerbib\@empty
\bibitem [{\citenamefont {Bednorz}\ and\ \citenamefont
  {Mueller}(1986)}]{1986Possible}%
  \BibitemOpen
  \bibfield  {author} {\bibinfo {author} {\bibfnamefont {J.~G.}\ \bibnamefont
  {Bednorz}}\ and\ \bibinfo {author} {\bibfnamefont {K.}~\bibnamefont
  {Mueller}},\ }\href
  {https://courses.physics.illinois.edu/phys596/fa2011/StudentWork/team3_final.pdf}
  {\bibfield  {journal} {\bibinfo  {journal} {Zeitschrift für Physik B
  Condensed Matter}\ }\textbf {\bibinfo {volume} {64}},\ \bibinfo {pages} {189}
  (\bibinfo {year} {1986})}\BibitemShut {NoStop}%
\bibitem [{\citenamefont {Lee}\ \emph {et~al.}(2006)\citenamefont {Lee},
  \citenamefont {Nagaosa},\ and\ \citenamefont {Wen}}]{RevModPhys.78.17}%
  \BibitemOpen
  \bibfield  {author} {\bibinfo {author} {\bibfnamefont {P.~A.}\ \bibnamefont
  {Lee}}, \bibinfo {author} {\bibfnamefont {N.}~\bibnamefont {Nagaosa}}, \ and\
  \bibinfo {author} {\bibfnamefont {X.-G.}\ \bibnamefont {Wen}},\ }\href
  {\doibase 10.1103/RevModPhys.78.17} {\bibfield  {journal} {\bibinfo
  {journal} {Rev. Mod. Phys.}\ }\textbf {\bibinfo {volume} {78}},\ \bibinfo
  {pages} {17} (\bibinfo {year} {2006})}\BibitemShut {NoStop}%
\bibitem [{\citenamefont {ANDERSON}(1987)}]{ANDERSON1196}%
  \BibitemOpen
  \bibfield  {author} {\bibinfo {author} {\bibfnamefont {P.~W.}\ \bibnamefont
  {ANDERSON}},\ }\href {\doibase 10.1126/science.235.4793.1196} {\bibfield
  {journal} {\bibinfo  {journal} {Science}\ }\textbf {\bibinfo {volume}
  {235}},\ \bibinfo {pages} {1196} (\bibinfo {year} {1987})}\BibitemShut
  {NoStop}%
\bibitem [{\citenamefont {Dagotto}(2005)}]{Dagotto2005}%
  \BibitemOpen
  \bibfield  {author} {\bibinfo {author} {\bibfnamefont {E.}~\bibnamefont
  {Dagotto}},\ }\href {\doibase 10.1126/science.1107559} {\bibfield  {journal}
  {\bibinfo  {journal} {Science}\ }\textbf {\bibinfo {volume} {309}},\ \bibinfo
  {pages} {257} (\bibinfo {year} {2005})}\BibitemShut {NoStop}%
\bibitem [{\citenamefont {Tranquada}\ \emph {et~al.}(1995)\citenamefont
  {Tranquada}, \citenamefont {Sternlieb}, \citenamefont {Axe}, \citenamefont
  {Nakamura},\ and\ \citenamefont {Uchida}}]{1995Evidence}%
  \BibitemOpen
  \bibfield  {author} {\bibinfo {author} {\bibfnamefont {J.~M.}\ \bibnamefont
  {Tranquada}}, \bibinfo {author} {\bibfnamefont {B.~J.}\ \bibnamefont
  {Sternlieb}}, \bibinfo {author} {\bibfnamefont {J.~D.}\ \bibnamefont {Axe}},
  \bibinfo {author} {\bibfnamefont {Y.}~\bibnamefont {Nakamura}}, \ and\
  \bibinfo {author} {\bibfnamefont {S.}~\bibnamefont {Uchida}},\ }\href
  {https://www.nature.com/articles/375561a0} {\bibfield  {journal} {\bibinfo
  {journal} {Nature}\ }\textbf {\bibinfo {volume} {375}},\ \bibinfo {pages}
  {561} (\bibinfo {year} {1995})}\BibitemShut {NoStop}%
\bibitem [{\citenamefont {Kivelson}\ \emph {et~al.}(2003)\citenamefont
  {Kivelson}, \citenamefont {Bindloss}, \citenamefont {Fradkin}, \citenamefont
  {Oganesyan}, \citenamefont {Tranquada}, \citenamefont {Kapitulnik},\ and\
  \citenamefont {Howald}}]{RevModPhys.75.1201}%
  \BibitemOpen
  \bibfield  {author} {\bibinfo {author} {\bibfnamefont {S.~A.}\ \bibnamefont
  {Kivelson}}, \bibinfo {author} {\bibfnamefont {I.~P.}\ \bibnamefont
  {Bindloss}}, \bibinfo {author} {\bibfnamefont {E.}~\bibnamefont {Fradkin}},
  \bibinfo {author} {\bibfnamefont {V.}~\bibnamefont {Oganesyan}}, \bibinfo
  {author} {\bibfnamefont {J.~M.}\ \bibnamefont {Tranquada}}, \bibinfo {author}
  {\bibfnamefont {A.}~\bibnamefont {Kapitulnik}}, \ and\ \bibinfo {author}
  {\bibfnamefont {C.}~\bibnamefont {Howald}},\ }\href {\doibase
  10.1103/RevModPhys.75.1201} {\bibfield  {journal} {\bibinfo  {journal} {Rev.
  Mod. Phys.}\ }\textbf {\bibinfo {volume} {75}},\ \bibinfo {pages} {1201}
  (\bibinfo {year} {2003})}\BibitemShut {NoStop}%
\bibitem [{\citenamefont {Zaanen}\ and\ \citenamefont
  {Gunnarsson}(1989)}]{PhysRevB.40.7391}%
  \BibitemOpen
  \bibfield  {author} {\bibinfo {author} {\bibfnamefont {J.}~\bibnamefont
  {Zaanen}}\ and\ \bibinfo {author} {\bibfnamefont {O.}~\bibnamefont
  {Gunnarsson}},\ }\href {\doibase 10.1103/PhysRevB.40.7391} {\bibfield
  {journal} {\bibinfo  {journal} {Phys. Rev. B}\ }\textbf {\bibinfo {volume}
  {40}},\ \bibinfo {pages} {7391} (\bibinfo {year} {1989})}\BibitemShut
  {NoStop}%
\bibitem [{\citenamefont {Machida}(1989)}]{MACHIDA1989192}%
  \BibitemOpen
  \bibfield  {author} {\bibinfo {author} {\bibfnamefont {K.}~\bibnamefont
  {Machida}},\ }\href {\doibase https://doi.org/10.1016/0921-4534(89)90316-X}
  {\bibfield  {journal} {\bibinfo  {journal} {Physica C: Superconductivity}\
  }\textbf {\bibinfo {volume} {158}},\ \bibinfo {pages} {192} (\bibinfo {year}
  {1989})}\BibitemShut {NoStop}%
\bibitem [{\citenamefont {Kato}\ \emph {et~al.}(1990)\citenamefont {Kato},
  \citenamefont {Machida}, \citenamefont {Nakanishi},\ and\ \citenamefont
  {Fujita}}]{doi:10.1143/JPSJ.59.1047}%
  \BibitemOpen
  \bibfield  {author} {\bibinfo {author} {\bibfnamefont {M.}~\bibnamefont
  {Kato}}, \bibinfo {author} {\bibfnamefont {K.}~\bibnamefont {Machida}},
  \bibinfo {author} {\bibfnamefont {H.}~\bibnamefont {Nakanishi}}, \ and\
  \bibinfo {author} {\bibfnamefont {M.}~\bibnamefont {Fujita}},\ }\href
  {\doibase 10.1143/JPSJ.59.1047} {\bibfield  {journal} {\bibinfo  {journal}
  {Journal of the Physical Society of Japan}\ }\textbf {\bibinfo {volume}
  {59}},\ \bibinfo {pages} {1047} (\bibinfo {year} {1990})}\BibitemShut
  {NoStop}%
\bibitem [{\citenamefont {Fujita}\ \emph {et~al.}(2002)\citenamefont {Fujita},
  \citenamefont {Goka}, \citenamefont {Yamada},\ and\ \citenamefont
  {Matsuda}}]{PhysRevLett.88.167008}%
  \BibitemOpen
  \bibfield  {author} {\bibinfo {author} {\bibfnamefont {M.}~\bibnamefont
  {Fujita}}, \bibinfo {author} {\bibfnamefont {H.}~\bibnamefont {Goka}},
  \bibinfo {author} {\bibfnamefont {K.}~\bibnamefont {Yamada}}, \ and\ \bibinfo
  {author} {\bibfnamefont {M.}~\bibnamefont {Matsuda}},\ }\href {\doibase
  10.1103/PhysRevLett.88.167008} {\bibfield  {journal} {\bibinfo  {journal}
  {Phys. Rev. Lett.}\ }\textbf {\bibinfo {volume} {88}},\ \bibinfo {pages}
  {167008} (\bibinfo {year} {2002})}\BibitemShut {NoStop}%
\bibitem [{\citenamefont {Abbamonte}\ \emph {et~al.}(2005)\citenamefont
  {Abbamonte}, \citenamefont {Rusydi}, \citenamefont {Smadici}, \citenamefont
  {Gu}, \citenamefont {Sawatzky},\ and\ \citenamefont {Feng}}]{2005Spatially}%
  \BibitemOpen
  \bibfield  {author} {\bibinfo {author} {\bibfnamefont {P.}~\bibnamefont
  {Abbamonte}}, \bibinfo {author} {\bibfnamefont {A.}~\bibnamefont {Rusydi}},
  \bibinfo {author} {\bibfnamefont {S.}~\bibnamefont {Smadici}}, \bibinfo
  {author} {\bibfnamefont {G.~D.}\ \bibnamefont {Gu}}, \bibinfo {author}
  {\bibfnamefont {G.~A.}\ \bibnamefont {Sawatzky}}, \ and\ \bibinfo {author}
  {\bibfnamefont {D.~L.}\ \bibnamefont {Feng}},\ }\href
  {https://www.nature.com/articles/nphys178} {\bibfield  {journal} {\bibinfo
  {journal} {Nature Physics}\ }\textbf {\bibinfo {volume} {1}},\ \bibinfo
  {pages} {155} (\bibinfo {year} {2005})}\BibitemShut {NoStop}%
\bibitem [{\citenamefont {Choubey}\ \emph {et~al.}(2020)\citenamefont
  {Choubey}, \citenamefont {Joo}, \citenamefont {Fujita}, \citenamefont {Du},
  \citenamefont {Edkins}, \citenamefont {Hamidian}, \citenamefont {Eisaki},
  \citenamefont {Uchida}, \citenamefont {Mackenzie},\ and\ \citenamefont
  {Lee}}]{2020Atomic}%
  \BibitemOpen
  \bibfield  {author} {\bibinfo {author} {\bibfnamefont {P.}~\bibnamefont
  {Choubey}}, \bibinfo {author} {\bibfnamefont {S.~H.}\ \bibnamefont {Joo}},
  \bibinfo {author} {\bibfnamefont {K.}~\bibnamefont {Fujita}}, \bibinfo
  {author} {\bibfnamefont {Z.}~\bibnamefont {Du}}, \bibinfo {author}
  {\bibfnamefont {S.~D.}\ \bibnamefont {Edkins}}, \bibinfo {author}
  {\bibfnamefont {M.~H.}\ \bibnamefont {Hamidian}}, \bibinfo {author}
  {\bibfnamefont {H.}~\bibnamefont {Eisaki}}, \bibinfo {author} {\bibfnamefont
  {S.}~\bibnamefont {Uchida}}, \bibinfo {author} {\bibfnamefont {A.~P.}\
  \bibnamefont {Mackenzie}}, \ and\ \bibinfo {author} {\bibfnamefont
  {J.}~\bibnamefont {Lee}},\ }\href
  {https://www.pnas.org/content/pnas/117/26/14805.full.pdf} {\bibfield
  {journal} {\bibinfo  {journal} {Proceedings of the National Academy of
  Sciences}\ }\textbf {\bibinfo {volume} {117}},\ \bibinfo {pages} {202002429}
  (\bibinfo {year} {2020})}\BibitemShut {NoStop}%
\bibitem [{\citenamefont {Zhu}\ \emph {et~al.}(2016)\citenamefont {Zhu},
  \citenamefont {Dong},\ and\ \citenamefont {Pu}}]{2016Chuanzhou}%
  \BibitemOpen
  \bibfield  {author} {\bibinfo {author} {\bibfnamefont {C.}~\bibnamefont
  {Zhu}}, \bibinfo {author} {\bibfnamefont {L.}~\bibnamefont {Dong}}, \ and\
  \bibinfo {author} {\bibfnamefont {H.}~\bibnamefont {Pu}},\ }\href {\doibase
  10.1088/0953-4075/49/14/145301} {\ \textbf {\bibinfo {volume} {49}},\
  \bibinfo {pages} {145301} (\bibinfo {year} {2016})}\BibitemShut {NoStop}%
\bibitem [{\citenamefont {Fradkin}\ \emph {et~al.}(2015)\citenamefont
  {Fradkin}, \citenamefont {Kivelson},\ and\ \citenamefont
  {Tranquada}}]{RevModPhys.87.457}%
  \BibitemOpen
  \bibfield  {author} {\bibinfo {author} {\bibfnamefont {E.}~\bibnamefont
  {Fradkin}}, \bibinfo {author} {\bibfnamefont {S.~A.}\ \bibnamefont
  {Kivelson}}, \ and\ \bibinfo {author} {\bibfnamefont {J.~M.}\ \bibnamefont
  {Tranquada}},\ }\href {\doibase 10.1103/RevModPhys.87.457} {\bibfield
  {journal} {\bibinfo  {journal} {Rev. Mod. Phys.}\ }\textbf {\bibinfo {volume}
  {87}},\ \bibinfo {pages} {457} (\bibinfo {year} {2015})}\BibitemShut
  {NoStop}%
\bibitem [{\citenamefont {Keimer}\ \emph {et~al.}(2015)\citenamefont {Keimer},
  \citenamefont {Kivelson}, \citenamefont {Norman}, \citenamefont {Uchida},\
  and\ \citenamefont {Zaanen}}]{nature14165}%
  \BibitemOpen
  \bibfield  {author} {\bibinfo {author} {\bibfnamefont {B.}~\bibnamefont
  {Keimer}}, \bibinfo {author} {\bibfnamefont {S.}~\bibnamefont {Kivelson}},
  \bibinfo {author} {\bibfnamefont {M.}~\bibnamefont {Norman}}, \bibinfo
  {author} {\bibfnamefont {S.}~\bibnamefont {Uchida}}, \ and\ \bibinfo {author}
  {\bibfnamefont {J.}~\bibnamefont {Zaanen}},\ }\href {\doibase
  10.1038/nature14165} {\bibfield  {journal} {\bibinfo  {journal} {Nature}\
  }\textbf {\bibinfo {volume} {518}},\ \bibinfo {pages} {179} (\bibinfo {year}
  {2015})}\BibitemShut {NoStop}%
\bibitem [{\citenamefont {Poilblanc}\ and\ \citenamefont
  {Rice}(1989)}]{PhysRevB.39.9749}%
  \BibitemOpen
  \bibfield  {author} {\bibinfo {author} {\bibfnamefont {D.}~\bibnamefont
  {Poilblanc}}\ and\ \bibinfo {author} {\bibfnamefont {T.~M.}\ \bibnamefont
  {Rice}},\ }\href {\doibase 10.1103/PhysRevB.39.9749} {\bibfield  {journal}
  {\bibinfo  {journal} {Phys. Rev. B}\ }\textbf {\bibinfo {volume} {39}},\
  \bibinfo {pages} {9749} (\bibinfo {year} {1989})}\BibitemShut {NoStop}%
\bibitem [{\citenamefont {Schulz}(1990)}]{PhysRevLett.64.1445}%
  \BibitemOpen
  \bibfield  {author} {\bibinfo {author} {\bibfnamefont {H.~J.}\ \bibnamefont
  {Schulz}},\ }\href {\doibase 10.1103/PhysRevLett.64.1445} {\bibfield
  {journal} {\bibinfo  {journal} {Phys. Rev. Lett.}\ }\textbf {\bibinfo
  {volume} {64}},\ \bibinfo {pages} {1445} (\bibinfo {year}
  {1990})}\BibitemShut {NoStop}%
\bibitem [{\citenamefont {Giamarchi}\ and\ \citenamefont
  {Lhuillier}(1990)}]{PhysRevB.42.10641}%
  \BibitemOpen
  \bibfield  {author} {\bibinfo {author} {\bibfnamefont {T.}~\bibnamefont
  {Giamarchi}}\ and\ \bibinfo {author} {\bibfnamefont {C.}~\bibnamefont
  {Lhuillier}},\ }\href {\doibase 10.1103/PhysRevB.42.10641} {\bibfield
  {journal} {\bibinfo  {journal} {Phys. Rev. B}\ }\textbf {\bibinfo {volume}
  {42}},\ \bibinfo {pages} {10641} (\bibinfo {year} {1990})}\BibitemShut
  {NoStop}%
\bibitem [{\citenamefont {Fye}\ \emph {et~al.}(1990)\citenamefont {Fye},
  \citenamefont {Martins},\ and\ \citenamefont {Scalettar}}]{PhysRevB.42.6809}%
  \BibitemOpen
  \bibfield  {author} {\bibinfo {author} {\bibfnamefont {R.~M.}\ \bibnamefont
  {Fye}}, \bibinfo {author} {\bibfnamefont {M.~J.}\ \bibnamefont {Martins}}, \
  and\ \bibinfo {author} {\bibfnamefont {R.~T.}\ \bibnamefont {Scalettar}},\
  }\href {\doibase 10.1103/PhysRevB.42.6809} {\bibfield  {journal} {\bibinfo
  {journal} {Phys. Rev. B}\ }\textbf {\bibinfo {volume} {42}},\ \bibinfo
  {pages} {6809} (\bibinfo {year} {1990})}\BibitemShut {NoStop}%
\bibitem [{\citenamefont {Moreo}\ \emph {et~al.}(1991)\citenamefont {Moreo},
  \citenamefont {Scalapino},\ and\ \citenamefont
  {Dagotto}}]{PhysRevB.43.11442}%
  \BibitemOpen
  \bibfield  {author} {\bibinfo {author} {\bibfnamefont {A.}~\bibnamefont
  {Moreo}}, \bibinfo {author} {\bibfnamefont {D.}~\bibnamefont {Scalapino}}, \
  and\ \bibinfo {author} {\bibfnamefont {E.}~\bibnamefont {Dagotto}},\ }\href
  {\doibase 10.1103/PhysRevB.43.11442} {\bibfield  {journal} {\bibinfo
  {journal} {Phys. Rev. B}\ }\textbf {\bibinfo {volume} {43}},\ \bibinfo
  {pages} {11442} (\bibinfo {year} {1991})}\BibitemShut {NoStop}%
\bibitem [{\citenamefont {Becca}\ \emph {et~al.}(2000)\citenamefont {Becca},
  \citenamefont {Capone},\ and\ \citenamefont {Sorella}}]{PhysRevB.62.12700}%
  \BibitemOpen
  \bibfield  {author} {\bibinfo {author} {\bibfnamefont {F.}~\bibnamefont
  {Becca}}, \bibinfo {author} {\bibfnamefont {M.}~\bibnamefont {Capone}}, \
  and\ \bibinfo {author} {\bibfnamefont {S.}~\bibnamefont {Sorella}},\ }\href
  {\doibase 10.1103/PhysRevB.62.12700} {\bibfield  {journal} {\bibinfo
  {journal} {Phys. Rev. B}\ }\textbf {\bibinfo {volume} {62}},\ \bibinfo
  {pages} {12700} (\bibinfo {year} {2000})}\BibitemShut {NoStop}%
\bibitem [{\citenamefont {Su}(1996)}]{PhysRevB.54.R8281}%
  \BibitemOpen
  \bibfield  {author} {\bibinfo {author} {\bibfnamefont {G.}~\bibnamefont
  {Su}},\ }\href {\doibase 10.1103/PhysRevB.54.R8281} {\bibfield  {journal}
  {\bibinfo  {journal} {Phys. Rev. B}\ }\textbf {\bibinfo {volume} {54}},\
  \bibinfo {pages} {R8281} (\bibinfo {year} {1996})}\BibitemShut {NoStop}%
\bibitem [{\citenamefont {Gehlhoff}(1996)}]{Gehlhoff_1996}%
  \BibitemOpen
  \bibfield  {author} {\bibinfo {author} {\bibfnamefont {L.}~\bibnamefont
  {Gehlhoff}},\ }\href {\doibase 10.1088/0953-8984/8/16/014} {\bibfield
  {journal} {\bibinfo  {journal} {Journal of Physics: Condensed Matter}\
  }\textbf {\bibinfo {volume} {8}},\ \bibinfo {pages} {2851} (\bibinfo {year}
  {1996})}\BibitemShut {NoStop}%
\bibitem [{\citenamefont {Emery}\ \emph {et~al.}(1990)\citenamefont {Emery},
  \citenamefont {Kivelson},\ and\ \citenamefont {Lin}}]{PhysRevLett.64.475}%
  \BibitemOpen
  \bibfield  {author} {\bibinfo {author} {\bibfnamefont {V.~J.}\ \bibnamefont
  {Emery}}, \bibinfo {author} {\bibfnamefont {S.~A.}\ \bibnamefont {Kivelson}},
  \ and\ \bibinfo {author} {\bibfnamefont {H.~Q.}\ \bibnamefont {Lin}},\ }\href
  {\doibase 10.1103/PhysRevLett.64.475} {\bibfield  {journal} {\bibinfo
  {journal} {Phys. Rev. Lett.}\ }\textbf {\bibinfo {volume} {64}},\ \bibinfo
  {pages} {475} (\bibinfo {year} {1990})}\BibitemShut {NoStop}%
\bibitem [{\citenamefont {Hellberg}\ and\ \citenamefont
  {Manousakis}(1997)}]{PhysRevLett.78.4609}%
  \BibitemOpen
  \bibfield  {author} {\bibinfo {author} {\bibfnamefont {C.~S.}\ \bibnamefont
  {Hellberg}}\ and\ \bibinfo {author} {\bibfnamefont {E.}~\bibnamefont
  {Manousakis}},\ }\href {\doibase 10.1103/PhysRevLett.78.4609} {\bibfield
  {journal} {\bibinfo  {journal} {Phys. Rev. Lett.}\ }\textbf {\bibinfo
  {volume} {78}},\ \bibinfo {pages} {4609} (\bibinfo {year}
  {1997})}\BibitemShut {NoStop}%
\bibitem [{\citenamefont {Gimm}\ and\ \citenamefont
  {Suck~Salk}(2000)}]{PhysRevB.62.13930}%
  \BibitemOpen
  \bibfield  {author} {\bibinfo {author} {\bibfnamefont {T.-H.}\ \bibnamefont
  {Gimm}}\ and\ \bibinfo {author} {\bibfnamefont {S.-H.}\ \bibnamefont
  {Suck~Salk}},\ }\href {\doibase 10.1103/PhysRevB.62.13930} {\bibfield
  {journal} {\bibinfo  {journal} {Phys. Rev. B}\ }\textbf {\bibinfo {volume}
  {62}},\ \bibinfo {pages} {13930} (\bibinfo {year} {2000})}\BibitemShut
  {NoStop}%
\bibitem [{\citenamefont {Putikka}\ and\ \citenamefont
  {Luchini}(2000)}]{PhysRevB.62.1684}%
  \BibitemOpen
  \bibfield  {author} {\bibinfo {author} {\bibfnamefont {W.~O.}\ \bibnamefont
  {Putikka}}\ and\ \bibinfo {author} {\bibfnamefont {M.~U.}\ \bibnamefont
  {Luchini}},\ }\href {\doibase 10.1103/PhysRevB.62.1684} {\bibfield  {journal}
  {\bibinfo  {journal} {Phys. Rev. B}\ }\textbf {\bibinfo {volume} {62}},\
  \bibinfo {pages} {1684} (\bibinfo {year} {2000})}\BibitemShut {NoStop}%
\bibitem [{\citenamefont {Shih}\ \emph {et~al.}(1998)\citenamefont {Shih},
  \citenamefont {Chen},\ and\ \citenamefont {Lee}}]{PhysRevB.57.627}%
  \BibitemOpen
  \bibfield  {author} {\bibinfo {author} {\bibfnamefont {C.~T.}\ \bibnamefont
  {Shih}}, \bibinfo {author} {\bibfnamefont {Y.~C.}\ \bibnamefont {Chen}}, \
  and\ \bibinfo {author} {\bibfnamefont {T.~K.}\ \bibnamefont {Lee}},\ }\href
  {\doibase 10.1103/PhysRevB.57.627} {\bibfield  {journal} {\bibinfo  {journal}
  {Phys. Rev. B}\ }\textbf {\bibinfo {volume} {57}},\ \bibinfo {pages} {627}
  (\bibinfo {year} {1998})}\BibitemShut {NoStop}%
\bibitem [{\citenamefont {Martins}\ \emph {et~al.}(2000)\citenamefont
  {Martins}, \citenamefont {Xavier}, \citenamefont {Gazza}, \citenamefont
  {Vojta},\ and\ \citenamefont {Dagotto}}]{PhysRevB.63.014414}%
  \BibitemOpen
  \bibfield  {author} {\bibinfo {author} {\bibfnamefont {G.~B.}\ \bibnamefont
  {Martins}}, \bibinfo {author} {\bibfnamefont {J.~C.}\ \bibnamefont {Xavier}},
  \bibinfo {author} {\bibfnamefont {C.}~\bibnamefont {Gazza}}, \bibinfo
  {author} {\bibfnamefont {M.}~\bibnamefont {Vojta}}, \ and\ \bibinfo {author}
  {\bibfnamefont {E.}~\bibnamefont {Dagotto}},\ }\href {\doibase
  10.1103/PhysRevB.63.014414} {\bibfield  {journal} {\bibinfo  {journal} {Phys.
  Rev. B}\ }\textbf {\bibinfo {volume} {63}},\ \bibinfo {pages} {014414}
  (\bibinfo {year} {2000})}\BibitemShut {NoStop}%
\bibitem [{\citenamefont {Chen}\ \emph {et~al.}(2021)\citenamefont {Chen},
  \citenamefont {Wang}, \citenamefont {Rebec}, \citenamefont {Jia},
  \citenamefont {Hashimoto}, \citenamefont {Lu}, \citenamefont {Moritz},
  \citenamefont {Moore}, \citenamefont {Devereaux},\ and\ \citenamefont
  {Shen}}]{doi:10.1126/science.abf5174}%
  \BibitemOpen
  \bibfield  {author} {\bibinfo {author} {\bibfnamefont {Z.}~\bibnamefont
  {Chen}}, \bibinfo {author} {\bibfnamefont {Y.}~\bibnamefont {Wang}}, \bibinfo
  {author} {\bibfnamefont {S.~N.}\ \bibnamefont {Rebec}}, \bibinfo {author}
  {\bibfnamefont {T.}~\bibnamefont {Jia}}, \bibinfo {author} {\bibfnamefont
  {M.}~\bibnamefont {Hashimoto}}, \bibinfo {author} {\bibfnamefont
  {D.}~\bibnamefont {Lu}}, \bibinfo {author} {\bibfnamefont {B.}~\bibnamefont
  {Moritz}}, \bibinfo {author} {\bibfnamefont {R.~G.}\ \bibnamefont {Moore}},
  \bibinfo {author} {\bibfnamefont {T.~P.}\ \bibnamefont {Devereaux}}, \ and\
  \bibinfo {author} {\bibfnamefont {Z.-X.}\ \bibnamefont {Shen}},\ }\href
  {\doibase 10.1126/science.abf5174} {\bibfield  {journal} {\bibinfo  {journal}
  {Science}\ }\textbf {\bibinfo {volume} {373}},\ \bibinfo {pages} {1235}
  (\bibinfo {year} {2021})}\BibitemShut {NoStop}%
\bibitem [{\citenamefont {Tranquada}\ \emph {et~al.}(1999)\citenamefont
  {Tranquada}, \citenamefont {Ichikawa},\ and\ \citenamefont
  {Uchida}}]{PhysRevB.59.14712}%
  \BibitemOpen
  \bibfield  {author} {\bibinfo {author} {\bibfnamefont {J.~M.}\ \bibnamefont
  {Tranquada}}, \bibinfo {author} {\bibfnamefont {N.}~\bibnamefont {Ichikawa}},
  \ and\ \bibinfo {author} {\bibfnamefont {S.}~\bibnamefont {Uchida}},\ }\href
  {\doibase 10.1103/PhysRevB.59.14712} {\bibfield  {journal} {\bibinfo
  {journal} {Phys. Rev. B}\ }\textbf {\bibinfo {volume} {59}},\ \bibinfo
  {pages} {14712} (\bibinfo {year} {1999})}\BibitemShut {NoStop}%
\bibitem [{\citenamefont {Lee}\ \emph {et~al.}(1999)\citenamefont {Lee},
  \citenamefont {Birgeneau}, \citenamefont {Kastner}, \citenamefont {Endoh},
  \citenamefont {Wakimoto}, \citenamefont {Yamada}, \citenamefont {Erwin},
  \citenamefont {Lee},\ and\ \citenamefont {Shirane}}]{PhysRevB.60.3643}%
  \BibitemOpen
  \bibfield  {author} {\bibinfo {author} {\bibfnamefont {Y.~S.}\ \bibnamefont
  {Lee}}, \bibinfo {author} {\bibfnamefont {R.~J.}\ \bibnamefont {Birgeneau}},
  \bibinfo {author} {\bibfnamefont {M.~A.}\ \bibnamefont {Kastner}}, \bibinfo
  {author} {\bibfnamefont {Y.}~\bibnamefont {Endoh}}, \bibinfo {author}
  {\bibfnamefont {S.}~\bibnamefont {Wakimoto}}, \bibinfo {author}
  {\bibfnamefont {K.}~\bibnamefont {Yamada}}, \bibinfo {author} {\bibfnamefont
  {R.~W.}\ \bibnamefont {Erwin}}, \bibinfo {author} {\bibfnamefont {S.-H.}\
  \bibnamefont {Lee}}, \ and\ \bibinfo {author} {\bibfnamefont
  {G.}~\bibnamefont {Shirane}},\ }\href {\doibase 10.1103/PhysRevB.60.3643}
  {\bibfield  {journal} {\bibinfo  {journal} {Phys. Rev. B}\ }\textbf {\bibinfo
  {volume} {60}},\ \bibinfo {pages} {3643} (\bibinfo {year}
  {1999})}\BibitemShut {NoStop}%
\bibitem [{\citenamefont {Zhong}\ \emph {et~al.}(2017)\citenamefont {Zhong},
  \citenamefont {Winn}, \citenamefont {Gu}, \citenamefont {Reznik},\ and\
  \citenamefont {Tranquada}}]{PhysRevLett.118.177601}%
  \BibitemOpen
  \bibfield  {author} {\bibinfo {author} {\bibfnamefont {R.}~\bibnamefont
  {Zhong}}, \bibinfo {author} {\bibfnamefont {B.~L.}\ \bibnamefont {Winn}},
  \bibinfo {author} {\bibfnamefont {G.}~\bibnamefont {Gu}}, \bibinfo {author}
  {\bibfnamefont {D.}~\bibnamefont {Reznik}}, \ and\ \bibinfo {author}
  {\bibfnamefont {J.~M.}\ \bibnamefont {Tranquada}},\ }\href {\doibase
  10.1103/PhysRevLett.118.177601} {\bibfield  {journal} {\bibinfo  {journal}
  {Phys. Rev. Lett.}\ }\textbf {\bibinfo {volume} {118}},\ \bibinfo {pages}
  {177601} (\bibinfo {year} {2017})}\BibitemShut {NoStop}%
\bibitem [{\citenamefont {Cheong}\ \emph {et~al.}(1991)\citenamefont {Cheong},
  \citenamefont {Aeppli}, \citenamefont {Mason}, \citenamefont {Mook},
  \citenamefont {Hayden}, \citenamefont {Canfield}, \citenamefont {Fisk},
  \citenamefont {Clausen},\ and\ \citenamefont
  {Martinez}}]{PhysRevLett.67.1791}%
  \BibitemOpen
  \bibfield  {author} {\bibinfo {author} {\bibfnamefont {S.-W.}\ \bibnamefont
  {Cheong}}, \bibinfo {author} {\bibfnamefont {G.}~\bibnamefont {Aeppli}},
  \bibinfo {author} {\bibfnamefont {T.~E.}\ \bibnamefont {Mason}}, \bibinfo
  {author} {\bibfnamefont {H.}~\bibnamefont {Mook}}, \bibinfo {author}
  {\bibfnamefont {S.~M.}\ \bibnamefont {Hayden}}, \bibinfo {author}
  {\bibfnamefont {P.~C.}\ \bibnamefont {Canfield}}, \bibinfo {author}
  {\bibfnamefont {Z.}~\bibnamefont {Fisk}}, \bibinfo {author} {\bibfnamefont
  {K.~N.}\ \bibnamefont {Clausen}}, \ and\ \bibinfo {author} {\bibfnamefont
  {J.~L.}\ \bibnamefont {Martinez}},\ }\href {\doibase
  10.1103/PhysRevLett.67.1791} {\bibfield  {journal} {\bibinfo  {journal}
  {Phys. Rev. Lett.}\ }\textbf {\bibinfo {volume} {67}},\ \bibinfo {pages}
  {1791} (\bibinfo {year} {1991})}\BibitemShut {NoStop}%
\bibitem [{\citenamefont {Yamada}\ \emph {et~al.}(1998)\citenamefont {Yamada},
  \citenamefont {Lee}, \citenamefont {Kurahashi}, \citenamefont {Wada},
  \citenamefont {Wakimoto}, \citenamefont {Ueki}, \citenamefont {Kimura},
  \citenamefont {Endoh}, \citenamefont {Hosoya}, \citenamefont {Shirane},
  \citenamefont {Birgeneau}, \citenamefont {Greven}, \citenamefont {Kastner},\
  and\ \citenamefont {Kim}}]{PhysRevB.57.6165}%
  \BibitemOpen
  \bibfield  {author} {\bibinfo {author} {\bibfnamefont {K.}~\bibnamefont
  {Yamada}}, \bibinfo {author} {\bibfnamefont {C.~H.}\ \bibnamefont {Lee}},
  \bibinfo {author} {\bibfnamefont {K.}~\bibnamefont {Kurahashi}}, \bibinfo
  {author} {\bibfnamefont {J.}~\bibnamefont {Wada}}, \bibinfo {author}
  {\bibfnamefont {S.}~\bibnamefont {Wakimoto}}, \bibinfo {author}
  {\bibfnamefont {S.}~\bibnamefont {Ueki}}, \bibinfo {author} {\bibfnamefont
  {H.}~\bibnamefont {Kimura}}, \bibinfo {author} {\bibfnamefont
  {Y.}~\bibnamefont {Endoh}}, \bibinfo {author} {\bibfnamefont
  {S.}~\bibnamefont {Hosoya}}, \bibinfo {author} {\bibfnamefont
  {G.}~\bibnamefont {Shirane}}, \bibinfo {author} {\bibfnamefont {R.~J.}\
  \bibnamefont {Birgeneau}}, \bibinfo {author} {\bibfnamefont {M.}~\bibnamefont
  {Greven}}, \bibinfo {author} {\bibfnamefont {M.~A.}\ \bibnamefont {Kastner}},
  \ and\ \bibinfo {author} {\bibfnamefont {Y.~J.}\ \bibnamefont {Kim}},\ }\href
  {\doibase 10.1103/PhysRevB.57.6165} {\bibfield  {journal} {\bibinfo
  {journal} {Phys. Rev. B}\ }\textbf {\bibinfo {volume} {57}},\ \bibinfo
  {pages} {6165} (\bibinfo {year} {1998})}\BibitemShut {NoStop}%
\bibitem [{\citenamefont {Mitrano}\ \emph {et~al.}(2019)\citenamefont
  {Mitrano}, \citenamefont {Lee}, \citenamefont {Husain}, \citenamefont
  {Delacretaz}, \citenamefont {Zhu}, \citenamefont {de~la Peña~Munoz},
  \citenamefont {Sun}, \citenamefont {Joe}, \citenamefont {Reid}, \citenamefont
  {Wandel}, \citenamefont {Coslovich}, \citenamefont {Schlotter}, \citenamefont
  {van Driel}, \citenamefont {Schneeloch}, \citenamefont {Gu}, \citenamefont
  {Hartnoll}, \citenamefont {Goldenfeld},\ and\ \citenamefont
  {Abbamonte}}]{doi:10.1126/sciadv.aax3346}%
  \BibitemOpen
  \bibfield  {author} {\bibinfo {author} {\bibfnamefont {M.}~\bibnamefont
  {Mitrano}}, \bibinfo {author} {\bibfnamefont {S.}~\bibnamefont {Lee}},
  \bibinfo {author} {\bibfnamefont {A.~A.}\ \bibnamefont {Husain}}, \bibinfo
  {author} {\bibfnamefont {L.}~\bibnamefont {Delacretaz}}, \bibinfo {author}
  {\bibfnamefont {M.}~\bibnamefont {Zhu}}, \bibinfo {author} {\bibfnamefont
  {G.}~\bibnamefont {de~la Peña~Munoz}}, \bibinfo {author} {\bibfnamefont
  {S.~X.-L.}\ \bibnamefont {Sun}}, \bibinfo {author} {\bibfnamefont {Y.~I.}\
  \bibnamefont {Joe}}, \bibinfo {author} {\bibfnamefont {A.~H.}\ \bibnamefont
  {Reid}}, \bibinfo {author} {\bibfnamefont {S.~F.}\ \bibnamefont {Wandel}},
  \bibinfo {author} {\bibfnamefont {G.}~\bibnamefont {Coslovich}}, \bibinfo
  {author} {\bibfnamefont {W.}~\bibnamefont {Schlotter}}, \bibinfo {author}
  {\bibfnamefont {T.}~\bibnamefont {van Driel}}, \bibinfo {author}
  {\bibfnamefont {J.}~\bibnamefont {Schneeloch}}, \bibinfo {author}
  {\bibfnamefont {G.~D.}\ \bibnamefont {Gu}}, \bibinfo {author} {\bibfnamefont
  {S.}~\bibnamefont {Hartnoll}}, \bibinfo {author} {\bibfnamefont
  {N.}~\bibnamefont {Goldenfeld}}, \ and\ \bibinfo {author} {\bibfnamefont
  {P.}~\bibnamefont {Abbamonte}},\ }\href {\doibase 10.1126/sciadv.aax3346}
  {\bibfield  {journal} {\bibinfo  {journal} {Science Advances}\ }\textbf
  {\bibinfo {volume} {5}},\ \bibinfo {pages} {eaax3346} (\bibinfo {year}
  {2019})}\BibitemShut {NoStop}%
\bibitem [{\citenamefont {Berg}\ \emph {et~al.}(2007)\citenamefont {Berg},
  \citenamefont {Fradkin}, \citenamefont {Kim}, \citenamefont {Kivelson},
  \citenamefont {Oganesyan}, \citenamefont {Tranquada},\ and\ \citenamefont
  {Zhang}}]{PhysRevLett.99.127003}%
  \BibitemOpen
  \bibfield  {author} {\bibinfo {author} {\bibfnamefont {E.}~\bibnamefont
  {Berg}}, \bibinfo {author} {\bibfnamefont {E.}~\bibnamefont {Fradkin}},
  \bibinfo {author} {\bibfnamefont {E.-A.}\ \bibnamefont {Kim}}, \bibinfo
  {author} {\bibfnamefont {S.~A.}\ \bibnamefont {Kivelson}}, \bibinfo {author}
  {\bibfnamefont {V.}~\bibnamefont {Oganesyan}}, \bibinfo {author}
  {\bibfnamefont {J.~M.}\ \bibnamefont {Tranquada}}, \ and\ \bibinfo {author}
  {\bibfnamefont {S.~C.}\ \bibnamefont {Zhang}},\ }\href {\doibase
  10.1103/PhysRevLett.99.127003} {\bibfield  {journal} {\bibinfo  {journal}
  {Phys. Rev. Lett.}\ }\textbf {\bibinfo {volume} {99}},\ \bibinfo {pages}
  {127003} (\bibinfo {year} {2007})}\BibitemShut {NoStop}%
\bibitem [{\citenamefont {Khatami}\ \emph {et~al.}(2015)\citenamefont
  {Khatami}, \citenamefont {Scalettar},\ and\ \citenamefont
  {Singh}}]{PhysRevB.91.241107}%
  \BibitemOpen
  \bibfield  {author} {\bibinfo {author} {\bibfnamefont {E.}~\bibnamefont
  {Khatami}}, \bibinfo {author} {\bibfnamefont {R.~T.}\ \bibnamefont
  {Scalettar}}, \ and\ \bibinfo {author} {\bibfnamefont {R.~R.~P.}\
  \bibnamefont {Singh}},\ }\href {\doibase 10.1103/PhysRevB.91.241107}
  {\bibfield  {journal} {\bibinfo  {journal} {Phys. Rev. B}\ }\textbf {\bibinfo
  {volume} {91}},\ \bibinfo {pages} {241107} (\bibinfo {year}
  {2015})}\BibitemShut {NoStop}%
\bibitem [{\citenamefont {White}\ \emph {et~al.}(1989)\citenamefont {White},
  \citenamefont {Scalapino}, \citenamefont {Sugar}, \citenamefont {Bickers},\
  and\ \citenamefont {Scalettar}}]{PhysRevB.39.839}%
  \BibitemOpen
  \bibfield  {author} {\bibinfo {author} {\bibfnamefont {S.~R.}\ \bibnamefont
  {White}}, \bibinfo {author} {\bibfnamefont {D.~J.}\ \bibnamefont
  {Scalapino}}, \bibinfo {author} {\bibfnamefont {R.~L.}\ \bibnamefont
  {Sugar}}, \bibinfo {author} {\bibfnamefont {N.~E.}\ \bibnamefont {Bickers}},
  \ and\ \bibinfo {author} {\bibfnamefont {R.~T.}\ \bibnamefont {Scalettar}},\
  }\href {\doibase 10.1103/PhysRevB.39.839} {\bibfield  {journal} {\bibinfo
  {journal} {Phys. Rev. B}\ }\textbf {\bibinfo {volume} {39}},\ \bibinfo
  {pages} {839} (\bibinfo {year} {1989})}\BibitemShut {NoStop}%
\bibitem [{\citenamefont {Hirsch}\ and\ \citenamefont
  {Lin}(1988)}]{PhysRevB.37.5070}%
  \BibitemOpen
  \bibfield  {author} {\bibinfo {author} {\bibfnamefont {J.~E.}\ \bibnamefont
  {Hirsch}}\ and\ \bibinfo {author} {\bibfnamefont {H.~Q.}\ \bibnamefont
  {Lin}},\ }\href {\doibase 10.1103/PhysRevB.37.5070} {\bibfield  {journal}
  {\bibinfo  {journal} {Phys. Rev. B}\ }\textbf {\bibinfo {volume} {37}},\
  \bibinfo {pages} {5070} (\bibinfo {year} {1988})}\BibitemShut {NoStop}%
\bibitem [{\citenamefont {Dagotto}(1994)}]{RevModPhys.66.763}%
  \BibitemOpen
  \bibfield  {author} {\bibinfo {author} {\bibfnamefont {E.}~\bibnamefont
  {Dagotto}},\ }\href {\doibase 10.1103/RevModPhys.66.763} {\bibfield
  {journal} {\bibinfo  {journal} {Rev. Mod. Phys.}\ }\textbf {\bibinfo {volume}
  {66}},\ \bibinfo {pages} {763} (\bibinfo {year} {1994})}\BibitemShut
  {NoStop}%
\bibitem [{\citenamefont {Ying}\ \emph {et~al.}(2014)\citenamefont {Ying},
  \citenamefont {Mondaini}, \citenamefont {Sun}, \citenamefont {Paiva},
  \citenamefont {Fye},\ and\ \citenamefont {Scalettar}}]{PhysRevB.90.075121}%
  \BibitemOpen
  \bibfield  {author} {\bibinfo {author} {\bibfnamefont {T.}~\bibnamefont
  {Ying}}, \bibinfo {author} {\bibfnamefont {R.}~\bibnamefont {Mondaini}},
  \bibinfo {author} {\bibfnamefont {X.~D.}\ \bibnamefont {Sun}}, \bibinfo
  {author} {\bibfnamefont {T.}~\bibnamefont {Paiva}}, \bibinfo {author}
  {\bibfnamefont {R.~M.}\ \bibnamefont {Fye}}, \ and\ \bibinfo {author}
  {\bibfnamefont {R.~T.}\ \bibnamefont {Scalettar}},\ }\href {\doibase
  10.1103/PhysRevB.90.075121} {\bibfield  {journal} {\bibinfo  {journal} {Phys.
  Rev. B}\ }\textbf {\bibinfo {volume} {90}},\ \bibinfo {pages} {075121}
  (\bibinfo {year} {2014})}\BibitemShut {NoStop}%
\bibitem [{\citenamefont {Zhang}\ \emph {et~al.}(1997)\citenamefont {Zhang},
  \citenamefont {Carlson},\ and\ \citenamefont
  {Gubernatis}}]{PhysRevB.55.7464}%
  \BibitemOpen
  \bibfield  {author} {\bibinfo {author} {\bibfnamefont {S.}~\bibnamefont
  {Zhang}}, \bibinfo {author} {\bibfnamefont {J.}~\bibnamefont {Carlson}}, \
  and\ \bibinfo {author} {\bibfnamefont {J.~E.}\ \bibnamefont {Gubernatis}},\
  }\href {\doibase 10.1103/PhysRevB.55.7464} {\bibfield  {journal} {\bibinfo
  {journal} {Phys. Rev. B}\ }\textbf {\bibinfo {volume} {55}},\ \bibinfo
  {pages} {7464} (\bibinfo {year} {1997})}\BibitemShut {NoStop}%
\bibitem [{\citenamefont {Zhang}\ \emph {et~al.}(1995)\citenamefont {Zhang},
  \citenamefont {Carlson},\ and\ \citenamefont
  {Gubernatis}}]{PhysRevLett.74.3652}%
  \BibitemOpen
  \bibfield  {author} {\bibinfo {author} {\bibfnamefont {S.}~\bibnamefont
  {Zhang}}, \bibinfo {author} {\bibfnamefont {J.}~\bibnamefont {Carlson}}, \
  and\ \bibinfo {author} {\bibfnamefont {J.~E.}\ \bibnamefont {Gubernatis}},\
  }\href {\doibase 10.1103/PhysRevLett.74.3652} {\bibfield  {journal} {\bibinfo
   {journal} {Phys. Rev. Lett.}\ }\textbf {\bibinfo {volume} {74}},\ \bibinfo
  {pages} {3652} (\bibinfo {year} {1995})}\BibitemShut {NoStop}%
\bibitem [{\citenamefont {Ma}\ \emph {et~al.}(2011)\citenamefont {Ma},
  \citenamefont {Huang}, \citenamefont {Hu},\ and\ \citenamefont
  {Lin}}]{PhysRevB.84.121410}%
  \BibitemOpen
  \bibfield  {author} {\bibinfo {author} {\bibfnamefont {T.}~\bibnamefont
  {Ma}}, \bibinfo {author} {\bibfnamefont {Z.}~\bibnamefont {Huang}}, \bibinfo
  {author} {\bibfnamefont {F.}~\bibnamefont {Hu}}, \ and\ \bibinfo {author}
  {\bibfnamefont {H.-Q.}\ \bibnamefont {Lin}},\ }\href {\doibase
  10.1103/PhysRevB.84.121410} {\bibfield  {journal} {\bibinfo  {journal} {Phys.
  Rev. B}\ }\textbf {\bibinfo {volume} {84}},\ \bibinfo {pages} {121410}
  (\bibinfo {year} {2011})}\BibitemShut {NoStop}%
\bibitem [{\citenamefont {Huang}\ \emph {et~al.}(2019)\citenamefont {Huang},
  \citenamefont {Zhang},\ and\ \citenamefont {Ma}}]{HUANG2019310}%
  \BibitemOpen
  \bibfield  {author} {\bibinfo {author} {\bibfnamefont {T.}~\bibnamefont
  {Huang}}, \bibinfo {author} {\bibfnamefont {L.}~\bibnamefont {Zhang}}, \ and\
  \bibinfo {author} {\bibfnamefont {T.}~\bibnamefont {Ma}},\ }\href {\doibase
  https://doi.org/10.1016/j.scib.2019.01.026} {\bibfield  {journal} {\bibinfo
  {journal} {Science Bulletin}\ }\textbf {\bibinfo {volume} {64}},\ \bibinfo
  {pages} {310 } (\bibinfo {year} {2019})}\BibitemShut {NoStop}%
\bibitem [{\citenamefont {Chen}\ \emph {et~al.}(2020)\citenamefont {Chen},
  \citenamefont {Chu}, \citenamefont {Huang},\ and\ \citenamefont
  {Ma}}]{PhysRevB.101.155413}%
  \BibitemOpen
  \bibfield  {author} {\bibinfo {author} {\bibfnamefont {W.}~\bibnamefont
  {Chen}}, \bibinfo {author} {\bibfnamefont {Y.}~\bibnamefont {Chu}}, \bibinfo
  {author} {\bibfnamefont {T.}~\bibnamefont {Huang}}, \ and\ \bibinfo {author}
  {\bibfnamefont {T.}~\bibnamefont {Ma}},\ }\href {\doibase
  10.1103/PhysRevB.101.155413} {\bibfield  {journal} {\bibinfo  {journal}
  {Phys. Rev. B}\ }\textbf {\bibinfo {volume} {101}},\ \bibinfo {pages}
  {155413} (\bibinfo {year} {2020})}\BibitemShut {NoStop}%
\bibitem [{\citenamefont {Loh}\ \emph {et~al.}(1990)\citenamefont {Loh},
  \citenamefont {Gubernatis}, \citenamefont {Scalettar}, \citenamefont {White},
  \citenamefont {Scalapino},\ and\ \citenamefont {Sugar}}]{PhysRevB.41.9301}%
  \BibitemOpen
  \bibfield  {author} {\bibinfo {author} {\bibfnamefont {E.~Y.}\ \bibnamefont
  {Loh}}, \bibinfo {author} {\bibfnamefont {J.~E.}\ \bibnamefont {Gubernatis}},
  \bibinfo {author} {\bibfnamefont {R.~T.}\ \bibnamefont {Scalettar}}, \bibinfo
  {author} {\bibfnamefont {S.~R.}\ \bibnamefont {White}}, \bibinfo {author}
  {\bibfnamefont {D.~J.}\ \bibnamefont {Scalapino}}, \ and\ \bibinfo {author}
  {\bibfnamefont {R.~L.}\ \bibnamefont {Sugar}},\ }\href {\doibase
  10.1103/PhysRevB.41.9301} {\bibfield  {journal} {\bibinfo  {journal} {Phys.
  Rev. B}\ }\textbf {\bibinfo {volume} {41}},\ \bibinfo {pages} {9301}
  (\bibinfo {year} {1990})}\BibitemShut {NoStop}%
\bibitem [{\citenamefont {Mondaini}\ \emph {et~al.}(2012)\citenamefont
  {Mondaini}, \citenamefont {Ying}, \citenamefont {Paiva},\ and\ \citenamefont
  {Scalettar}}]{PhysRevB.86.184506}%
  \BibitemOpen
  \bibfield  {author} {\bibinfo {author} {\bibfnamefont {R.}~\bibnamefont
  {Mondaini}}, \bibinfo {author} {\bibfnamefont {T.}~\bibnamefont {Ying}},
  \bibinfo {author} {\bibfnamefont {T.}~\bibnamefont {Paiva}}, \ and\ \bibinfo
  {author} {\bibfnamefont {R.~T.}\ \bibnamefont {Scalettar}},\ }\href {\doibase
  10.1103/PhysRevB.86.184506} {\bibfield  {journal} {\bibinfo  {journal} {Phys.
  Rev. B}\ }\textbf {\bibinfo {volume} {86}},\ \bibinfo {pages} {184506}
  (\bibinfo {year} {2012})}\BibitemShut {NoStop}%
\bibitem [{\citenamefont {Li}\ \emph {et~al.}()\citenamefont {Li},
  \citenamefont {Tian}, \citenamefont {Liang},\ and\ \citenamefont
  {Ma}}]{li2021dopingdependent}%
  \BibitemOpen
  \bibfield  {author} {\bibinfo {author} {\bibfnamefont {Y.}~\bibnamefont
  {Li}}, \bibinfo {author} {\bibfnamefont {L.}~\bibnamefont {Tian}}, \bibinfo
  {author} {\bibfnamefont {Y.}~\bibnamefont {Liang}}, \ and\ \bibinfo {author}
  {\bibfnamefont {T.}~\bibnamefont {Ma}},\ }\href@noop {} {}\Eprint
  {http://arxiv.org/abs/2106.03296} {arXiv:2106.03296} \BibitemShut {NoStop}%
\bibitem [{\citenamefont {Tranquada}(1995)}]{1995Stripe}%
  \BibitemOpen
  \bibfield  {author} {\bibinfo {author} {\bibfnamefont {J.~M.}\ \bibnamefont
  {Tranquada}},\ }\href {https://www.nature.com/articles/375561a0} {\bibfield
  {journal} {\bibinfo  {journal} {Nature}\ }\textbf {\bibinfo {volume} {375}},\
  \bibinfo {pages} {561} (\bibinfo {year} {1995})}\BibitemShut {NoStop}%
\bibitem [{\citenamefont {Tranquada}\ \emph {et~al.}(1997)\citenamefont
  {Tranquada}, \citenamefont {Axe}, \citenamefont {Ichikawa}, \citenamefont
  {Moodenbaugh}, \citenamefont {Nakamura},\ and\ \citenamefont
  {Uchida}}]{PhysRevLett.78.338}%
  \BibitemOpen
  \bibfield  {author} {\bibinfo {author} {\bibfnamefont {J.~M.}\ \bibnamefont
  {Tranquada}}, \bibinfo {author} {\bibfnamefont {J.~D.}\ \bibnamefont {Axe}},
  \bibinfo {author} {\bibfnamefont {N.}~\bibnamefont {Ichikawa}}, \bibinfo
  {author} {\bibfnamefont {A.~R.}\ \bibnamefont {Moodenbaugh}}, \bibinfo
  {author} {\bibfnamefont {Y.}~\bibnamefont {Nakamura}}, \ and\ \bibinfo
  {author} {\bibfnamefont {S.}~\bibnamefont {Uchida}},\ }\href {\doibase
  10.1103/PhysRevLett.78.338} {\bibfield  {journal} {\bibinfo  {journal} {Phys.
  Rev. Lett.}\ }\textbf {\bibinfo {volume} {78}},\ \bibinfo {pages} {338}
  (\bibinfo {year} {1997})}\BibitemShut {NoStop}%
\bibitem [{\citenamefont {Wilson}\ \emph {et~al.}(1975)\citenamefont {Wilson},
  \citenamefont {Salvo},\ and\ \citenamefont {Mahajan}}]{1975Charge}%
  \BibitemOpen
  \bibfield  {author} {\bibinfo {author} {\bibfnamefont {J.}~\bibnamefont
  {Wilson}}, \bibinfo {author} {\bibfnamefont {F.~D.}\ \bibnamefont {Salvo}}, \
  and\ \bibinfo {author} {\bibfnamefont {S.}~\bibnamefont {Mahajan}},\ }\href
  {\doibase 10.1080/00018737500101391} {\bibfield  {journal} {\bibinfo
  {journal} {Advances in Physics}\ }\textbf {\bibinfo {volume} {24}},\ \bibinfo
  {pages} {117} (\bibinfo {year} {1975})}\BibitemShut {NoStop}%
\bibitem [{\citenamefont {Gibbs}\ \emph {et~al.}(1988)\citenamefont {Gibbs},
  \citenamefont {Mohanty},\ and\ \citenamefont {Bohr}}]{PhysRevB.37.562}%
  \BibitemOpen
  \bibfield  {author} {\bibinfo {author} {\bibfnamefont {D.}~\bibnamefont
  {Gibbs}}, \bibinfo {author} {\bibfnamefont {K.~M.}\ \bibnamefont {Mohanty}},
  \ and\ \bibinfo {author} {\bibfnamefont {J.}~\bibnamefont {Bohr}},\ }\href
  {\doibase 10.1103/PhysRevB.37.562} {\bibfield  {journal} {\bibinfo  {journal}
  {Phys. Rev. B}\ }\textbf {\bibinfo {volume} {37}},\ \bibinfo {pages} {562}
  (\bibinfo {year} {1988})}\BibitemShut {NoStop}%
\bibitem [{\citenamefont {Monceau}(2012)}]{Pierre2012}%
  \BibitemOpen
  \bibfield  {author} {\bibinfo {author} {\bibfnamefont {P.}~\bibnamefont
  {Monceau}},\ }\href {\doibase 10.1080/00018732.2012.719674} {\bibfield
  {journal} {\bibinfo  {journal} {Advances in Physics}\ }\textbf {\bibinfo
  {volume} {61}},\ \bibinfo {pages} {325} (\bibinfo {year} {2012})}\BibitemShut
  {NoStop}%
\bibitem [{\citenamefont {Ghiringhelli}\ \emph {et~al.}(2012)\citenamefont
  {Ghiringhelli}, \citenamefont {Le~Tacon}, \citenamefont {Minola},
  \citenamefont {Blanco-Canosa}, \citenamefont {Mazzoli}, \citenamefont
  {Brookes}, \citenamefont {De~Luca}, \citenamefont {Frano}, \citenamefont
  {Hawthorn}, \citenamefont {He}, \citenamefont {Loew}, \citenamefont {Sala},
  \citenamefont {Peets}, \citenamefont {Salluzzo}, \citenamefont {Schierle},
  \citenamefont {Sutarto}, \citenamefont {Sawatzky}, \citenamefont {Weschke},
  \citenamefont {Keimer},\ and\ \citenamefont {Braicovich}}]{Ghiringhelli821}%
  \BibitemOpen
  \bibfield  {author} {\bibinfo {author} {\bibfnamefont {G.}~\bibnamefont
  {Ghiringhelli}}, \bibinfo {author} {\bibfnamefont {M.}~\bibnamefont
  {Le~Tacon}}, \bibinfo {author} {\bibfnamefont {M.}~\bibnamefont {Minola}},
  \bibinfo {author} {\bibfnamefont {S.}~\bibnamefont {Blanco-Canosa}}, \bibinfo
  {author} {\bibfnamefont {C.}~\bibnamefont {Mazzoli}}, \bibinfo {author}
  {\bibfnamefont {N.~B.}\ \bibnamefont {Brookes}}, \bibinfo {author}
  {\bibfnamefont {G.~M.}\ \bibnamefont {De~Luca}}, \bibinfo {author}
  {\bibfnamefont {A.}~\bibnamefont {Frano}}, \bibinfo {author} {\bibfnamefont
  {D.~G.}\ \bibnamefont {Hawthorn}}, \bibinfo {author} {\bibfnamefont
  {F.}~\bibnamefont {He}}, \bibinfo {author} {\bibfnamefont {T.}~\bibnamefont
  {Loew}}, \bibinfo {author} {\bibfnamefont {M.~M.}\ \bibnamefont {Sala}},
  \bibinfo {author} {\bibfnamefont {D.~C.}\ \bibnamefont {Peets}}, \bibinfo
  {author} {\bibfnamefont {M.}~\bibnamefont {Salluzzo}}, \bibinfo {author}
  {\bibfnamefont {E.}~\bibnamefont {Schierle}}, \bibinfo {author}
  {\bibfnamefont {R.}~\bibnamefont {Sutarto}}, \bibinfo {author} {\bibfnamefont
  {G.~A.}\ \bibnamefont {Sawatzky}}, \bibinfo {author} {\bibfnamefont
  {E.}~\bibnamefont {Weschke}}, \bibinfo {author} {\bibfnamefont
  {B.}~\bibnamefont {Keimer}}, \ and\ \bibinfo {author} {\bibfnamefont
  {L.}~\bibnamefont {Braicovich}},\ }\href {\doibase 10.1126/science.1223532}
  {\bibfield  {journal} {\bibinfo  {journal} {Science}\ }\textbf {\bibinfo
  {volume} {337}},\ \bibinfo {pages} {821} (\bibinfo {year}
  {2012})}\BibitemShut {NoStop}%
\bibitem [{\citenamefont {Blanco-Canosa}\ \emph {et~al.}(2014)\citenamefont
  {Blanco-Canosa}, \citenamefont {Frano}, \citenamefont {Schierle},
  \citenamefont {Porras}, \citenamefont {Loew}, \citenamefont {Minola},
  \citenamefont {Bluschke}, \citenamefont {Weschke}, \citenamefont {Keimer},\
  and\ \citenamefont {Le~Tacon}}]{PhysRevB.90.054513}%
  \BibitemOpen
  \bibfield  {author} {\bibinfo {author} {\bibfnamefont {S.}~\bibnamefont
  {Blanco-Canosa}}, \bibinfo {author} {\bibfnamefont {A.}~\bibnamefont
  {Frano}}, \bibinfo {author} {\bibfnamefont {E.}~\bibnamefont {Schierle}},
  \bibinfo {author} {\bibfnamefont {J.}~\bibnamefont {Porras}}, \bibinfo
  {author} {\bibfnamefont {T.}~\bibnamefont {Loew}}, \bibinfo {author}
  {\bibfnamefont {M.}~\bibnamefont {Minola}}, \bibinfo {author} {\bibfnamefont
  {M.}~\bibnamefont {Bluschke}}, \bibinfo {author} {\bibfnamefont
  {E.}~\bibnamefont {Weschke}}, \bibinfo {author} {\bibfnamefont
  {B.}~\bibnamefont {Keimer}}, \ and\ \bibinfo {author} {\bibfnamefont
  {M.}~\bibnamefont {Le~Tacon}},\ }\href {\doibase 10.1103/PhysRevB.90.054513}
  {\bibfield  {journal} {\bibinfo  {journal} {Phys. Rev. B}\ }\textbf {\bibinfo
  {volume} {90}},\ \bibinfo {pages} {054513} (\bibinfo {year}
  {2014})}\BibitemShut {NoStop}%
\bibitem [{\citenamefont {He}\ \emph {et~al.}(2014)\citenamefont {He},
  \citenamefont {Yin}, \citenamefont {Zech}, \citenamefont {Soumyanarayanan},
  \citenamefont {Yee}, \citenamefont {Williams}, \citenamefont {Boyer},
  \citenamefont {Chatterjee}, \citenamefont {Wise}, \citenamefont {Zeljkovic},
  \citenamefont {Kondo}, \citenamefont {Takeuchi}, \citenamefont {Ikuta},
  \citenamefont {Mistark}, \citenamefont {Markiewicz}, \citenamefont {Bansil},
  \citenamefont {Sachdev}, \citenamefont {Hudson},\ and\ \citenamefont
  {Hoffman}}]{He608}%
  \BibitemOpen
  \bibfield  {author} {\bibinfo {author} {\bibfnamefont {Y.}~\bibnamefont
  {He}}, \bibinfo {author} {\bibfnamefont {Y.}~\bibnamefont {Yin}}, \bibinfo
  {author} {\bibfnamefont {M.}~\bibnamefont {Zech}}, \bibinfo {author}
  {\bibfnamefont {A.}~\bibnamefont {Soumyanarayanan}}, \bibinfo {author}
  {\bibfnamefont {M.~M.}\ \bibnamefont {Yee}}, \bibinfo {author} {\bibfnamefont
  {T.}~\bibnamefont {Williams}}, \bibinfo {author} {\bibfnamefont {M.~C.}\
  \bibnamefont {Boyer}}, \bibinfo {author} {\bibfnamefont {K.}~\bibnamefont
  {Chatterjee}}, \bibinfo {author} {\bibfnamefont {W.~D.}\ \bibnamefont
  {Wise}}, \bibinfo {author} {\bibfnamefont {I.}~\bibnamefont {Zeljkovic}},
  \bibinfo {author} {\bibfnamefont {T.}~\bibnamefont {Kondo}}, \bibinfo
  {author} {\bibfnamefont {T.}~\bibnamefont {Takeuchi}}, \bibinfo {author}
  {\bibfnamefont {H.}~\bibnamefont {Ikuta}}, \bibinfo {author} {\bibfnamefont
  {P.}~\bibnamefont {Mistark}}, \bibinfo {author} {\bibfnamefont {R.~S.}\
  \bibnamefont {Markiewicz}}, \bibinfo {author} {\bibfnamefont
  {A.}~\bibnamefont {Bansil}}, \bibinfo {author} {\bibfnamefont
  {S.}~\bibnamefont {Sachdev}}, \bibinfo {author} {\bibfnamefont {E.~W.}\
  \bibnamefont {Hudson}}, \ and\ \bibinfo {author} {\bibfnamefont {J.~E.}\
  \bibnamefont {Hoffman}},\ }\href {\doibase 10.1126/science.1248221}
  {\bibfield  {journal} {\bibinfo  {journal} {Science}\ }\textbf {\bibinfo
  {volume} {344}},\ \bibinfo {pages} {608} (\bibinfo {year}
  {2014})}\BibitemShut {NoStop}%
\bibitem [{\citenamefont {Zhang}\ \emph {et~al.}(2000)\citenamefont {Zhang},
  \citenamefont {Carlson},\ and\ \citenamefont
  {Gubernatis}}]{PhysRevLett.84.2550}%
  \BibitemOpen
  \bibfield  {author} {\bibinfo {author} {\bibfnamefont {S.}~\bibnamefont
  {Zhang}}, \bibinfo {author} {\bibfnamefont {J.}~\bibnamefont {Carlson}}, \
  and\ \bibinfo {author} {\bibfnamefont {J.~E.}\ \bibnamefont {Gubernatis}},\
  }\href {\doibase 10.1103/PhysRevLett.84.2550} {\bibfield  {journal} {\bibinfo
   {journal} {Phys. Rev. Lett.}\ }\textbf {\bibinfo {volume} {84}},\ \bibinfo
  {pages} {2550} (\bibinfo {year} {2000})}\BibitemShut {NoStop}%
\bibitem [{\citenamefont {Huang}\ \emph {et~al.}(2001)\citenamefont {Huang},
  \citenamefont {Lin},\ and\ \citenamefont {Gubernatis}}]{PhysRevB.64.205101}%
  \BibitemOpen
  \bibfield  {author} {\bibinfo {author} {\bibfnamefont {Z.~B.}\ \bibnamefont
  {Huang}}, \bibinfo {author} {\bibfnamefont {H.~Q.}\ \bibnamefont {Lin}}, \
  and\ \bibinfo {author} {\bibfnamefont {J.~E.}\ \bibnamefont {Gubernatis}},\
  }\href {\doibase 10.1103/PhysRevB.64.205101} {\bibfield  {journal} {\bibinfo
  {journal} {Phys. Rev. B}\ }\textbf {\bibinfo {volume} {64}},\ \bibinfo
  {pages} {205101} (\bibinfo {year} {2001})}\BibitemShut {NoStop}%
\end{thebibliography}%

\end{document}